\documentclass[12pt]{article}

\usepackage[T1]{fontenc}
\usepackage{float}
\usepackage{amsmath}
\usepackage{graphicx}
\usepackage{setspace}
\usepackage[authoryear]{natbib}
\usepackage[latin9]{inputenc}
\usepackage{lmodern}

\usepackage{algpseudocode}
\usepackage{algorithm}
\usepackage{algorithmicx}

\usepackage{subfig}
\usepackage{amsmath,amssymb,amsthm,setspace,paralist}
\usepackage[colorlinks,citecolor=blue,urlcolor=blue]{hyperref}
\usepackage[normalem]{ulem}

\newcommand{\bsb}{\boldsymbol}

\newcommand{\bsby}{\boldsymbol{y}}
\newcommand{\bsbb}{\boldsymbol{\beta}}

\newcommand{\bsbY}{\boldsymbol{Y}}
\newcommand{\bsbB}{\boldsymbol{B}}
\newcommand{\bsbA}{\boldsymbol{A}}

\newcommand{\bsbV}{\boldsymbol{V}}
\newcommand{\bsbU}{\boldsymbol{U}}

\newcommand{\bsbE}{\boldsymbol{E}}
\newcommand{\bsbX}{\boldsymbol{X}}
\newcommand{\bsbZ}{\boldsymbol{Z}}

\newcommand{\bsbu}{\boldsymbol{u}}
\newcommand{\bsbI}{\boldsymbol{I}}
\newcommand{\bsbS}{\boldsymbol{S}}

\newcommand{\bsbmu}{{\boldsymbol{\mu}}}
\newcommand{\bsbx}{\boldsymbol{x}}
\newcommand{\bsbT}{\boldsymbol{T}}
\newcommand{\bsbW}{\boldsymbol{W}}

\newcommand{\bsbv}{\boldsymbol{v}}
\newcommand{\bsbz}{\boldsymbol{z}}

\newcommand{\bsbP}{\boldsymbol{P}}
\newcommand{\bsbQ}{\boldsymbol{Q}}
\newcommand{\bsbL}{\boldsymbol{L}}

\newcommand{\bsbg}{\boldsymbol{\gamma}}

\newcommand{\bsbalpha}{\boldsymbol{\alpha}}

\DeclareMathOperator{\sgn}{sgn}
\DeclareMathOperator{\vect}{\mbox{vec}\,}

\newcommand{\rd}{\,\mathrm{d}}

\theoremstyle{definition}
\newtheorem{example}{\protect\examplename}
\theoremstyle{plain}
\providecommand{\examplename}{Example}

\newtheorem{remark}{Remark}

\usepackage{xr}
\begin{document}

\title{On Generalization and Computation of Tukey's Depth: Part II}
\author{Yiyuan She, Shao Tang, and Jingze Liu\\Department of Statistics, Florida State University}%
\date{}
\maketitle


\begin{abstract}
This paper studies how to generalize Tukey's  depth to problems  defined in a restricted space that may be curved or have boundaries, and to problems with a nondifferentiable  objective. First, using a manifold approach, we  propose a broad class of Riemannian
depth for smooth problems defined on a Riemannian manifold, and  showcase its applications in spherical data analysis, principal component analysis, and multivariate orthogonal regression. Moreover, for nonsmooth problems, we introduce additional slack variables and inequality constraints to define a novel slacked data depth, which can perform center-outward rankings of estimators arising from  sparse learning and reduced rank regression. Real data examples illustrate the usefulness of some proposed data depths.

\end{abstract}

\section{Introduction}
\label{sec:Extending-Depth-with}
Tukey's half-space depth \citep{tukey1975mathematics} can be  generalized to a polished subspace depth, as shown in our companion paper \citep{SheDepthI2021}.
The basic Tukeyfication process there
assumes a simple problem structure in the sense that one can directly write down
some sample-additive estimating equations. Modern statistical applications however
pose
new
challenges.

First,   the parameter space $\Omega$ can be   curved or have boundaries, so that  evaluating the   gradient    in the ambient Euclidean space may {not} directly deliver reasonable influences. Second, with a  regularizer in use, the objective function is typically non-differentiable. Some specific examples are given as follows.
\begin{example}
\textbf{(Watson depth)} \label{exa:(Spherical-data-depth)} Assume
that all data points   lie on an $m$-dimensional sphere    $\boldsymbol{z}_{i}\in\mathbb{S}^{m-1}$,
and $\pm\boldsymbol{z}_{i} $ are deemed equivalent.
 This kind of data are typically referred to as axially symmetric data.  They  have recently received  attention
in clustering and directional statistics \citep{dhillon2003,bijral2007,SRA2013}.
To characterize the distribution of  such data,
a commonly used one is the Watson distribution with    density \citep{Watson1965,mardia1999directional}
  $$p(\boldsymbol{z}; \boldsymbol{\mu},\kappa )\propto e^{\kappa(\boldsymbol{\mu}^{T}\boldsymbol{z})^{2}}.$$ Here, $\boldsymbol{\mu}\in\mathbb{S}^{m-1}$ gives the mean direction,
$\kappa $ is the so-called concentration parameter, and the normalizing
constant does not depend on $\boldsymbol{\mu}$. We require $\kappa\neq0$ (otherwise  $\boldsymbol{\mu}$ is not an effective parameter to introduce
   depth). When $\kappa>0$, the data points concentrate
around $\boldsymbol{\mu}$,   and when $\kappa<0$,
the data spread around the great circle orthogonal to $\boldsymbol{\mu}$.
How to ``Tukeyfy'' more complex distributions defined on a sphere (such as the \textit{Fisher-Bingham  distribution}) is nontrivial, but could give rise to more useful
spherical data depths. We will see that depth-enhanced principal component analysis to be introduced in Section \ref{subsec:depthpca} poses a similar manifold challenge.
\end{example}
\begin{example}
\textbf{(Nonnegative regression depth)} \label{exa:(Non-negative-regression-data)}
As an extension of the celebrated regression depth \citep{Rousseeuw1999regression}, let us
consider data depth in a
  setting where all coefficients are nonnegative. This corresponds
to the nonnegative least squares problem: $$\min_{\boldsymbol{\beta}}\lVert\boldsymbol{y}-\boldsymbol{X}\boldsymbol{\beta}\rVert_{2}^{2}\;\textrm{s.t.}\;\beta_{j}\geq0, 1\leq j\leq p ,$$ where we can denote the constraints by $\boldsymbol{\beta}\in \mathbb R_+^p$ with $\mathbb R_+  = [0, \infty)$. Clearly, the closed parameter space $\mathbb R_+^p$ has boundary points.   Regression depth can be simply applied if    $\boldsymbol{\beta}^{\circ}$
is an interior point, but if $\boldsymbol{\beta}^{\circ}$ lies on the
boundary, i.e., $\beta_{j}^{\circ}=0$ for some $j$, which is of
 practical interest in significance tests, regular   depth does not apply, and the normal-equation based influences must  be corrected---but how?
\end{example}
\begin{example}
\textbf{(Sparsity depth)} \label{exa:(Lasso-depth)} Consider a sparse learning problem    $$\min_{\bsbb}  \sum_{i=1}^n \allowbreak l_0(\bsbx_i^T \bsbb) + \sum_{j=1}^p P(|\beta_j|; \lambda ),$$ where $l_0$ is a loss function defined on the systematic component $\bsbx_i^T \bsbb$ and
$P$ is a
penalty function to promote   sparsity in $\bsbb$. Examples of $P$ include  $\ell_1$,  $\ell_0$,  SCAD \citep{Fan2001}, and MCP \citep{Zhang2010MCP},   among many others that are popularly used in high dimensional statistics   for building a parsimonious
model. We assume that the regularization parameter $\lambda$ is
given, either by theory---see, e.g., \cite{Cai2009},  or by tuning,
like cross-validation \citep{SheCV}, so that the criterion is fully specified. A new class of depths  like   $\ell_1$-depth or $\ell_0$-depth   would be helpful for   high-dimensional robust inference, but the    nondifferentiability and  nonconvexity    of $P$  make  it difficult to obtain sample-additive estimation equations.
\end{example}

To  tackle the challenges, we  propose two approaches based on manifolds and slack variables, respectively, to extend Tukey's depth to Riemannian depth and slacked depth.
The rest of the paper is organized as follows.
Section  \ref{subsec:rgrad} studies how to handle a smooth problem defined on a Riemannian manifold. The resulting Riemannian
depth  finds  applications in spherical data analysis, principal component analysis, and multivariate orthogonal regression. Section \ref{subsec:slack} uses slack variables to cope with
parameter spaces with boundaries and
 nondifferentiable objectives. A novel class of slacked data depth can perform center-outward rankings of estimators arising from sparse learning and reduced rank regression.
Section \ref{sec:experiments} performs computer experiments on some real data examples. We conclude the whole work in Section \ref{sec:summ}. 

\paragraph{Notation. }
We use bold  symbols to denote vectors and matrices. A matrix $\boldsymbol{X}\in\mathbb{R}^{n\times p}$
is frequently partitioned into rows $\boldsymbol{X}=[\boldsymbol{x}_{1}\ldots\boldsymbol{x}_{n}]^{T}$
with $\boldsymbol{x}_{i}\in\mathbb{R}^{p}$. The vectorization   of   $\boldsymbol{X}$ is denoted
by $\textrm{vec}(\boldsymbol{X})\in\mathbb{R}^{np}$.
Let $\mathbb R_+=[0, +\infty]$.  We use  $\bsbX[\mathcal I, \mathcal J]$ to  denote a submatrix of $\bsbX$ with  rows and columns indexed by $\mathcal I$ and $\mathcal J$, respectively, and occasionally abbreviate $\bsbX[,\mathcal J ]$ to $\bsbX_{\mathcal J}$ by selecting  the corresponding columns. Given   $\boldsymbol{X}\in\mathbb{R}^{n\times p}$, $\lVert\boldsymbol{X}\rVert_{F}$
and $\lVert\boldsymbol{X}\rVert_{2}$ denote its Frobenius norm and
spectral norm, respectively, $\| \bsbX\|_{\max}\triangleq  \max_{1\le i\le n, 1\le j\le p} |x_{ij}|$, and $\mbox{rank}(\bsbX)$ denotes its rank. The Moore-Penrose inverse of $\bsbX$ is denoted by $\bsbX^+$. The inner product of two matrices $\boldsymbol{X}$
and $\boldsymbol{Y}$ (of the same size) is defined as $\langle\boldsymbol{X},\boldsymbol{Y}\rangle=\textrm{Tr}(\boldsymbol{X}^{T}\boldsymbol{Y})$
and their element-wise product (Hadamard product) is $\boldsymbol{X}\circ\boldsymbol{Y}$.
  The Kronecker product is denoted by $\boldsymbol{X}\otimes\boldsymbol{Y}$
(where $\boldsymbol{X}$ and $\boldsymbol{Y}$ need not have the same
dimensions). Given  a set $\mathcal A \subset \mathbb R^{p\times m}$ and a matrix $\bsbT \in \mathbb R^{n\times p}$,    $\bsbT \circ \mathcal A = \{\bsbT \bsbA: \bsbA\in \mathcal A \}$. We use $\mathbb{O}^{m\times r}$ to represent  the set of all $m\times r$ matrices $\bsbV$ satisfying the orthogonality constraint $\bsbV^T\bsbV =\bsbI$.   For a vector $\bsb{a}=[a_1,\ldots,a_n]^{T}$ $\in \mathbb{R}^{n}$, $\mbox{diag}\{\bsb{a}\}$ is defined as an $n \times n$ diagonal matrix with diagonal entries given by $a_1,\ldots,a_n$, and for a square matrix $\bsb{A}=[a_{ij}]_{n\times n}$, $\mbox{diag}(\bsb{A}):=\mbox{diag}\{ a_{11},\ldots,a_{nn}\}$.  The indicator
function $1_{\mathcal{A}}(t)$ means $1_{\mathcal{A}}(t)=1$ if $t\in\mathcal{A}$ and
$0$ otherwise.
Given $f:\mathbb{R}^{n\times p}\rightarrow\mathbb{R}$,
  $f\in \mathcal C^1$ means that its Euclidean gradient  $\nabla f(\boldsymbol{X})$, an $n\times p$ matrix with the $(i,j)$ element
$ {\partial f }/{\partial x_{ij}}$,   exists and is continuous for any $\bsbX \in \mathbb R^{n\times p}$. Given two vectors $\bsb{\alpha}, \bsbb\in \mathbb R^p$, $\bsb{\alpha}\succeq \bsbb$ means $\alpha_j \ge \beta_j, 1\le  j \le p$ and $\bsb{\alpha}\succ \bsbb$ means $\alpha_j > \beta_j, 1\le  j \le p$.
Finally, $a\wedge b = \min\{a, b\}$.

\section{A Manifold Approach}
\label{subsec:rgrad}
When the problem is defined on   a Riemannian manifold (without boundaries), we can introduce Riemannian influences, along with defining a proper influence space to complete the definition of  Riemannian depth.   In contrast, the commonly used methods to deal with   constraints (such as the elimination approach   for   \eqref{or-11} below) may be  infeasible   in higher dimensions. We will see the important role of the influence space $\mathcal G$ introduced in Section \ref{sec:Tukeyfication} of \cite{SheDepthI2021}, since    a Riemannian gradient  always lies in a tangent space.
\subsection{Riemannian depth}
\label{subsec:RieDep}
We begin with   Example \ref{exa:(Spherical-data-depth)} to motivate the main idea. Starting from such an example, the Watson depth will be introduced. Useful for the analysis of axial data, it is defined on a Riemmanian manifold and it will be a special case of a more generic Riemannian depth.

 For the MLE problem
\begin{equation}
\min_{\boldsymbol{\mu}}-\kappa\sum_{i}\langle\boldsymbol{\mu},\boldsymbol{z}_{i}\rangle^{2} + c(\kappa; m) \;\textrm{s.t.}\;\lVert\boldsymbol{\mu}\rVert_{2}^{2}=1,\label{eq:MLE waston}
\end{equation}
Lagrange multiplier or eigenvalue decomposition can be used
to solve for $\boldsymbol{\mu}$, but they do not yield a simple set of estimating equations like  \eqref{eq:sample addtivity} in \cite{SheDepthI2021} to be conveniently used for the purpose of  data   depth.

Instead, we view (\ref{eq:MLE waston})
as an {\textit{unconstrained}} problem on the sphere $\mathbb{S}^{m-1}$,
which is a Riemannian manifold. Then  the Riemannian gradient with
respect to $\boldsymbol{\mu}$ can be calculated to define the desired (Riemannian) influence function. We show some detailed derivation to give the reader more intuition.

Concretely, adopting the {canonical} metric induced by the inner product $G_{\boldsymbol{\mu}}(\boldsymbol{u}_1,\boldsymbol{u}_2)=\boldsymbol{u}_1^{T}(\boldsymbol{I}-\boldsymbol{\mu}\boldsymbol{\mu}^{T}/2)\boldsymbol{u}_2$
\citep{Alan1998},   the Riemannian gradient of $l_{i} :=-\kappa\langle\boldsymbol{\mu},\boldsymbol{z}_{i}\rangle^{2}$
 with respect
to $\boldsymbol{\mu}$, denoted by $\boldsymbol{g}_{i}(\boldsymbol{\mu})$,
is  defined as the unique element in the tangent space $$\mathcal{T}_{\boldsymbol{\mu}}(\mathbb{S}^{m-1})\triangleq\{\boldsymbol{u}\in\mathbb{R}^{m}:\boldsymbol{u}^{T}\boldsymbol{\mu}=0\}$$
satisfying
\begin{align} \label{eq:Rgraddef}
G_{\boldsymbol{\mu}}(\boldsymbol{g}_{i}(\boldsymbol{\mu}),\boldsymbol{u})=\boldsymbol{u}^{T}\nabla l_{i}, \ \forall \boldsymbol{u}\in\mathcal{T}_{\boldsymbol{\mu}}(\mathbb{S}^{m-1})\end{align}
 where    $\nabla l_{i}$ is the    Euclidean gradient.   It follows that   $$\boldsymbol{g}_{i}(\boldsymbol{\mu})=[\nabla l_{i}\boldsymbol{\mu}^{T}-\boldsymbol{\mu}(\nabla l_{i})^{T}]\boldsymbol{\mu}=-2\kappa\langle\boldsymbol{z}_{i},\boldsymbol{\mu}\rangle(\boldsymbol{z}_{i}-\langle\boldsymbol{z}_{i},\boldsymbol{\mu}\rangle\boldsymbol{\mu}).$$
From \cite{boothby1986introduction} (and  $\kappa\neq0$),
the optimal $\bsbmu$ satisfies
\begin{equation}
\sum_{i}\langle\boldsymbol{z}_{i},\boldsymbol{\mu}\rangle(\boldsymbol{z}_{i}-\langle\boldsymbol{z}_{i},\boldsymbol{\mu}\rangle\boldsymbol{\mu})=\boldsymbol{0}.\label{eq:waston gradient}
\end{equation}
Given $\bsbmu^\circ \in \mathbb S^{m-1}$,  the Riemannian  influence  $\langle\boldsymbol{z}_{i},\boldsymbol{\mu}^\circ\rangle(\boldsymbol{z}_{i}-\langle\boldsymbol{z}_{i},\boldsymbol{\mu}^\circ\rangle\boldsymbol{\mu}^\circ)$, denoted by $\boldsymbol{T}^{\mbox{\tiny R}}(\boldsymbol{\mu}^{\circ}; \boldsymbol{z}_{i})$,
  is no longer  $\boldsymbol{z}_{i}- \boldsymbol{\mu}^\circ$  as in location  depth.  
 Notably,  $\boldsymbol{T}^{\mbox{\tiny R}}(\boldsymbol{\mu}^{\circ}; \boldsymbol{z}_{i})$ vanishes  when $\theta_{i}= j\pi/2$   $(j=0,1,2,3)$ with $\cos \theta_i = \langle \bsbz_i, \bsbmu^\circ\rangle$,
corresponding to various circumstances with
$\kappa>0$ and $\kappa<0$.


Not only does the manifold perspective  provide     the desirable estimation equations, but it defines an important  influence space $\mathcal G  = \mathcal{T}_{\boldsymbol{\mu}^{\circ}}(\mathbb{S}^{m-1})  $ to   {restrict} $\bsbv$. Accordingly, our Watson depth  considers all    one-dimensional
projections \textit{tangentially}  passing through   $\boldsymbol{\mu}^{\circ}$:
\begin{equation}
d_{01}^{\mbox{\tiny W}}(\boldsymbol{\mu}^{\circ})=\min_{\boldsymbol{v}}\sum_{i}1_{\ge 0}(G_{\boldsymbol{\mu}^{\circ}}( \boldsymbol{v},(\boldsymbol{z}_{i}^{T}\boldsymbol{\mu}^{\circ})[\boldsymbol{z}_{i}-(\boldsymbol{z}_{i}^{T}\boldsymbol{\mu}^{\circ})\boldsymbol{\mu}^{\circ}]))\;\textrm{s.t.}\;\boldsymbol{v}^{T}\boldsymbol{\mu}^{\circ}=0,\boldsymbol{v}^{T}\boldsymbol{v}=1 \end{equation}
or equivalently \begin{equation}
d_{01}^{\mbox{\tiny W}}(\boldsymbol{\mu}^{\circ})=\min_{\boldsymbol{v}}\sum_{i}1_{\ge 0}(\langle\boldsymbol{v},(\boldsymbol{z}_{i}^{T}\boldsymbol{\mu}^{\circ})\boldsymbol{z}_{i}\rangle)\;\textrm{s.t.}\;\boldsymbol{v}^{T}\boldsymbol{\mu}^{\circ}=0,\boldsymbol{v}^{T}\boldsymbol{v}=1,\label{eq:spherical depth}
\end{equation}
regardless of the Riemannian metric, as an outcome of \eqref{eq:Rgraddef}.
The factor $\boldsymbol{z}_{i}^{T}\boldsymbol{\mu}^{\circ}$ in \eqref{eq:spherical depth}, possibly   negative,  amounts to replacing $\bsbz_i$ by $\sgn \langle\boldsymbol{z}_{i}^{T},\boldsymbol{\mu}^{\circ}\rangle \cdot \bsbz_i$. This  is in accordance with the Watson distribution    for  axially symmetric spherical data. 
The algorithms
 in Section \ref{sec:Data-depth-computation}  of our companion paper can be applied, after a   simple reparametrization of $\bsbv$ in the orthogonal complement space of $\bsbmu^\circ \bsbmu^{\circ T}$.

The above derivation is standard and can be generalized to introduce  a \textit{Riemannian depth}  for  the Tukeyfication of  a differentiable  loss $l$   on a Riemannian manifold $\mathcal M$ of an Euclidean space: $\min_{\bsbB} \sum _i l(\bsbB; \bsbx_i, \bsby_i)$ s.t. $\bsbB\in \mathcal M$. Given a point $\bsbB^\circ\in \mathcal M$ of interest, letting $\bsbT_i^\circ=\bsbT(\bsbB^\circ; \bsbx_i, \bsby_i) = \nabla_{\bsbB} l(\bsbB^\circ; \bsbx_i, \bsby_i)$ as before and considering   all directional derivatives of $l$ in the  directions of     $\boldsymbol{V}\in  \mathcal T_{\bsbB^\circ} (\mathcal M)$,
 we  define
\begin{align}
\mbox{\textbf{Riemannian depth:}} \  \ d_{01}^{\mbox{\tiny R}}(\bsbB^\circ)  =  \ & \ \min_{\boldsymbol{V} }   \sum_{i}1_{\ge 0}(\langle\boldsymbol{V},\boldsymbol{T}_{i}^{\circ}\rangle) \notag\\&  \ \textrm{ s.t. }   \boldsymbol{V}\in\mathcal T_{\bsbB^\circ} (\mathcal M), \lVert\boldsymbol{V}\rVert_{F}=1. \label{eq:defRdepth}
\end{align}
 Eqn. \eqref{eq:defRdepth} performs location depth of Riemannian influences in the tangent space $\mathcal T_{\bsbB^\circ} (\mathcal M)$. Because  $\mathcal T_{\bsbB^\circ} (\mathcal M)$ is  linear, the restricted   Procrustes
rotation in   Section \ref{sec:Data-depth-computation} of \cite{SheDepthI2021} applies with no difficulty in  optimization.

 When  $\mathcal M$ is compact
and/or $l$ is nonconvex, it becomes necessary  to  exclude locally maximal solutions in  the estimating equations. We give an ``\textit{order-2 Tukeyfication}'' as follows. Given $\bsbB^\circ \in \mathcal M$ and $\boldsymbol{V}\in\mathcal T_{\bsbB^\circ} (\mathcal M)$, let $\gamma$  be the geodesic satisfying $\gamma(0)= \bsbB^\circ$ and $\gamma'(0) = \bsbV$. The  first step  is   to  restrict   $l$ to the geodesic and define
 $$
 g_i = \frac{\rd }{\rd t} l(\gamma(t); \bsbx_i, \bsby_i)\big\rvert_{t=0, }, \quad h_i = \frac{\rd^2 }{\rd t^2} l(\gamma(t); \bsbx_i, \bsby_i)\big\rvert_{t=0, }
 $$
 where $g_i$ simplifies  to $\langle\boldsymbol{V},\boldsymbol{T}_{i}^{\circ}\rangle$ and $h_i$ can be calculated via Riemannian Hessian. (Our companion paper mostly considers an Euclidean  $\mathcal M$, where a line restriction $l(\bsbB^\circ +t \bsbV)$ with $\bsbV\ne \bsb0$ is used, and   $g_i$ and $h_i$ only involve the ordinary  gradient and Hessian of  $l$.)
The second step robustly measures    how well the following two optimality condition are obeyed:
$$
\sum_{i=1}^n g_i = 0, \quad \sum_{i=1}^n h_i \ge 0.
$$
Concretely,  changing   the one-dimensional averages to  medians motivates us to     adopt  $\sum \left( 1_{=0} + 2 ( 1_{<0} \wedge 1_{>0})\right)  (g_i)$ and $\sum 1_{\ge 0} ( h_i)  $ to  quantify  to what extent the two conditions  are satisfied, respectively, in the possible occurrence of extreme outliers. Finally, combining the two measures leads to
\begin{align*}
\mbox{\textbf{Riemannian depth (order 2):}}   \ d_{01}^{\mbox{\tiny R2}}(\bsbB^\circ)  =  \ & \ \min_{\boldsymbol{V} }   \sum_{i}1_{\gtrsim  0}(g_i ) \sum _i 1_{\ge 0}(h_i) \notag\\&   \textrm{s.t. }   \boldsymbol{V}\in\mathcal T_{\bsbB^\circ} (\mathcal M), \lVert\boldsymbol{V}\rVert_{F}=1,
\end{align*}
where $1_{\gtrsim  0}   :=  0.5 \cdot 1_{=0} +     1_{>0}  $ replaces   $    1_{=0} + 2 ( 1_{<0} \wedge 1_{>0}) $ in the optimization because   $g_i$ is linear in $\bsbV$, the Riemannian Hessian is a bilinear map,  and $T_{\bsbB^\circ} (\mathcal M)$ is a linear space. A perhaps more aggressive but amenable proposal is to use $   \sum_{i}1_{\gtrsim  0}(g_i )   1_{\ge 0}(h_i)$ as the criterion.  (Notice the mild difference between  $1_{\gtrsim 0}$ and $1_{\ge 0}$; the first   seems to be more appropriate to deal with   equality-type optimality conditions  in defining a $d_{01}$-type data depth.)   When $l$ is (geodesically) convex, $h_i\ge 0$ and thus  the associated factor with proper scaling, $ \sum _i 1_{\ge 0}(h_i)/n$, will not affect the  depth.
 \begin{remark} \label{rmk:spdepth}
If we {Tukeyfy} the basic  \emph{von Mises-Fisher  distribution} (vMF) \citep{mardia1999directional}, with the density given by
  $p(\boldsymbol{z}; \boldsymbol{\mu},\kappa )\propto e^{\kappa \boldsymbol{\mu}^{T}\boldsymbol{z} }$, where $\bsbmu: \| \bsbmu\|_2=1$ is the mean direction,   $\kappa>0 $ and the normalizing constant does not depend on $\bsbmu$,   \eqref{eq:defRdepth} yields $
d_{01}^{\mbox{\tiny R}}(\boldsymbol{\mu}^{\circ})=\min_{\boldsymbol{v}}\sum_{i}1_{\ge 0}(\langle\boldsymbol{v},  \boldsymbol{z}_{i}\rangle)\;\textrm{s.t.}\;\boldsymbol{v}^{T}\boldsymbol{\mu}^{\circ}=0,\boldsymbol{v}^{T}\boldsymbol{v}=1. $
This is closely related to but different from the angular Tukey's depth that can be defined as  $\min_{\boldsymbol{v}}\sum_{i}1_{\ge 0}(\langle\boldsymbol{v},  \boldsymbol{z}_{i}\rangle)\;\textrm{s.t.}\;\boldsymbol{v}^{T}\boldsymbol{\mu}^{\circ}\ge 0,\boldsymbol{v}^{T}\boldsymbol{v}=1$ for $m$-dimensional spherical data  (see \cite{Regina1992} for some theoretical studies when $m=2,3$). The order-2 depths involve $h_i =\langle\boldsymbol{\mu}^\circ,  \boldsymbol{z}_{i}\rangle$  which are independent of $\bsbv$ in this case.

More interesting notions of spherical data depth can  be induced by some more flexible distributions through  our manifold framework, such as     the    {\emph{Kent distribution}}
and  the more  general   {\emph{Fisher-Bingham  distribution}}  whose  quadratic exponential form  is more powerful than vMF for  statistical modeling  in bioinformatics, meteorology, and computer vision. 
\end{remark}

\subsection{Depth-enhanced principal component analysis}
\label{subsec:depthpca}
This part uses the Riemannian depth introduced in the last subsection to Tukeyfy the well-known principal component analysis (PCA). Let $\boldsymbol{Z}=[\bsbz_i, \ldots,\allowbreak \bsbz_n]^T\in \mathbb R^{n\times m}$
be a   data matrix.
The PCA model can be stated as
\begin{align} \bsbZ   = \bsb1\bsbmu^{* T} + \bsbA^* \bsbU^{*T} + \bsbE,\label{pcaass}\end{align} with    $\bsbmu^*\in \mathbb R^{m}$, $\bsbA^*\in \mathbb R^{n\times r}$, and $\bsbU^*\in \mathbb O^{n\times r}$  all unknown. Eqn. \eqref{pcaass} means that  the $n$ data points, after some proper translation,  all approximately concentrate in an $r$-dimensional subspace, and  $r$ is typically much lower than $m$ and $n$. The columns of $\bsbU^*$ are  often called the principal component (\textbf{PC}) loading directions. Assuming  that the entries of $\bsbE$ are i.i.d. Gaussian, we can estimate the intercept vector and  the low-dimensional  subspace by
\begin{align}\min_{(\boldsymbol{U}, \,\boldsymbol{\mu})}\lVert (\boldsymbol{Z}-\boldsymbol{1}\boldsymbol{\mu}^{T})(\bsbI - \boldsymbol{U} \bsbU^T)\rVert_{F}^{2} \quad \textrm{s.t.}\quad \boldsymbol{U}^{T}\boldsymbol{U}=\boldsymbol{I}_{r\times r}. \label{pcaprob}\end{align}
The solution is given by standard PCA, which is however sensitive to outliers.

One may want to estimate  $ \bsbU, \bsbmu$ more robustly through a depth enhancement. Here, the orthogonality constraint $\bsbU ^T \bsbU =\bsbI$   may appear more complex than that in  the spherical problem \eqref{eq:MLE waston}, but \eqref{pcaprob}  is  a smooth problem  on  a Stiefel manifold. Therefore, we can define a Riemannian depth for any $(\bsbmu^\circ, \bsbU^\circ)\in \mathbb R^m\times \mathbb O^{m\times r}$ based on Section \ref{subsec:RieDep}, which we call the    \textit{principal component} (\textbf{PC})  depth, as follows
\begin{align}
\mbox{\textbf{PC-depth}:}\ \  \min_{( \boldsymbol{v}, \boldsymbol{V})   } &  \sum_{i}1_{\ge 0}(\langle\boldsymbol{v},(\bsbI - \bsbU^{\circ}\bsbU^{\circ T}) (\bsbmu^\circ - \bsbz_i) \rangle-\langle\boldsymbol{V},(\bsbmu^\circ - \bsbz_i)(\bsbmu^\circ - \bsbz_i)^T\bsbU^\circ\rangle) \notag \\
 & \textrm{ s.t. }  \boldsymbol{V}^{T}\boldsymbol{U}^{\circ}+ \boldsymbol{U}^{\circ T}  \boldsymbol{V}=\boldsymbol{0}, \ \|\boldsymbol{v}\|_2^2 + \|\boldsymbol{V}\|_F^2 = 1. 
\label{pcdepth}
\end{align}
 All matrix differentiation details are  omitted. \eqref{pcdepth} may need an order-2 modification though, which will be clearly revealed by comparing it to \eqref{multioregdepth} later. 
\\

 PCA is also helpful when ranking observations in ultra-high dimensions. It is well known that the curse of dimensionality may make every observation look like a corner point, thus harmful to describing data depth. Fortunately, under   \eqref{pcaass},  the true signals  concentrate in the PC subspace determined by $\bsbU^*$; so to check a given point's centrality or extremity, it is helpful to  project it  onto the orthogonal complement  (\textbf{OC}) subspace to  reveal its outlyingness. See  \cite{She2016b} for more discussions. Specifically,  letting  $\bar \bsbU^*\in \mathbb O^{m\times \bar r}$ ($\bar r\le m-r$) that is orthogonal to  $\bsbU^*$, we can obtain from    \eqref{pcaass}  \begin{align}\bsbZ   \bar \bsbU^*= \bsb1\bar\bsbmu^{* T} +   \bar \bsbE, \label{ocpcaass}\end{align} where $\bar \bsbmu^* = \bar \bsbU^{*T} \bsbmu^*$, $\bar \bsbE= \bsbE \bar \bsbU^*$.  Eqn. \eqref{ocpcaass} is in the typical  location estimation setting except that   $\bar \bsbU^*$ is  unknown, which motivates us to consider    \begin{align}\min_{(\bar{\boldsymbol{U}}, \,\bar{\boldsymbol{\mu}})}\lVert\boldsymbol{Z}\bar {\boldsymbol{U}}-\boldsymbol{1}\bar{\boldsymbol{\mu}}^{T}\rVert_{F}^{2} \quad \textrm{s.t.}\quad \bar{\boldsymbol{U}}^{T}\bar{\boldsymbol{U}}=\boldsymbol{I}_{\bar r\times \bar r}. \label{or-gen}\end{align}
 Interestingly, \eqref{or-gen} can also be viewed as a \emph{multivariate} extension,  of rank $\bar r$,  of the  orthogonal
regression due to    \cite{mizera2002}: \begin{align}\min_{ \mu \in \mathbb R, \bsbu\in \mathbb R^{p+1}} \left\| \, [\boldsymbol{X}\;\boldsymbol{y}]\, \bsbu - \bsb1 \,\mu \, \right \|_2^2 \mbox{ s.t. }\| \bsbu\|_2^2=1.\label{11oreg}\end{align} Moreover, when $\boldsymbol{Z}=[\boldsymbol{X}\;\boldsymbol{Y}]$,  setting  $\bar {\boldsymbol{U}} =[\boldsymbol{B}^{T}\;\boldsymbol{\Gamma}^{T}]^{T}$  gives a  model  $\boldsymbol{Y}\boldsymbol{\Gamma}+\boldsymbol{X}\boldsymbol{B}-\boldsymbol{1}\bar{\boldsymbol{\mu}}^{T}=\boldsymbol{E}$ for canonical correlation analysis.

How to introduce an operational depth  for    
\eqref{or-gen} is a meaningful   problem. Indeed,  with  a deep  $\bar \bsbU$ provided,  one would be able to rank  high-dimensional  samples in a lower dimensional subspace.

 Restricting    to a naive case of  \eqref{11oreg}  with
a single
predictor and  a single response:
\begin{align}\min_{(\alpha, \beta, \mu)}\lVert\beta \boldsymbol{x} + \alpha \boldsymbol{y}-\boldsymbol{1} {\mu} \rVert_{F}^{2}\;\textrm{s.t.}\; \alpha^2 + \beta^2 = 1, \label{or-11}\end{align}
 one  can \textit{eliminate} the constraint  by, say, $\alpha = -\sin t, \beta = \cos t$ with a    free parameter $t$,  and then take the Euclidean gradient  with respect to $(t, \mu)$ to define a tangent depth \citep{mizera2002}. Nevertheless,  the   elimination method encounters difficulties when considering  multiple predictors, let alone  a general $\bar r$. As far as we know, there exists  no    commonly acknowledged  multivariate orthogonal regression depth  in the literature.

Our manifold approach provides a systematic treatment of  \eqref{or-gen} for all $p$,   $m$, and $\bar r$. We  call the resulting Riemannian depth the \textit{orthogonal complement}  (\textbf{OC})  depth. It     pursues  an  $\bar r$-dimensional subspace  in the original input space to       rank the observations effectively. The   influence space     here is        $\mathbb{R}^{\bar r}\times\mathcal{T}_{\bar{\boldsymbol{U}}}(\mathbb{O}^{  m\times \bar r})\  \mbox{ with } \ \mathcal{T}_{\bar{\boldsymbol{U}} }(\mathbb{O}^{  m\times \bar r})= \{\boldsymbol{V}:\bar{\boldsymbol{U}}^{T}\boldsymbol{V} + \boldsymbol{V}^{T}\bar{\boldsymbol{U}}=\boldsymbol{0}\}, $ and the OC  depth for any given $(\bar{\boldsymbol{\mu}}^{\circ},\bar{\boldsymbol{U}}^{\circ})\in \mathbb R^{\bar r} \times  \mathbb{O}^{  m\times \bar r}$ is
{
\begin{align}
\begin{split}
\mbox{\textbf{OC-depth}:}\ \ \min_{( \boldsymbol{v}, \boldsymbol{V}) }  & \sum_{i}  1_{\ge 0}(\langle\boldsymbol{v},\bar{\boldsymbol{\mu}}^\circ-\bar{\boldsymbol{U}}^{\circ T}\boldsymbol{z}_{i}\rangle+\langle\boldsymbol{V}, \bsbz_i \bsbz_i^T \bar\bsbU^\circ - \bsbz_i\bar\bsbmu^{\circ T}\rangle) \\
& \textrm{ s.t. }    {\boldsymbol{V}}^{T}\bar{\boldsymbol{U}}^{\circ}+ \bar{\boldsymbol{U}}^{\circ T}  \boldsymbol{V}=\boldsymbol{0}, \ \|\boldsymbol{v}\|_2^2 + \|\boldsymbol{V}\|_F^2 = 1. \end{split}\label{multioregdepth}
\end{align}}
 The derivations are similar to the PC depth and are omitted. Note that the      influence space constraint  has a multivariate form but  is linear. Eqn. \eqref{multioregdepth} also gives a  multivariate   orthogonal regression depth.

On the other hand, with   $r=\bar r$ and no intercepts, the (order-1) PC-depth and OC-depth  coincide,  since  in this case the two losses in \eqref{pcaprob} and \eqref{or-gen}  only differ by a minus sign  and the influence space is linear in $\bsbV$. This means that the most  and  least informative subspaces will have the same depth. As aforementioned, an order-2 Riemannian depth would be able to  distinguish between  minimization and maximization problems,   which deserves further investigation.    

\section{The Slack Variable Approach}
\label{subsec:slack}
The  nondifferentiability issue in  the other two examples in Section  \ref{sec:Extending-Depth-with} is much trickier to cope with.
In more detail, Example \ref{exa:(Non-negative-regression-data)}
has a closed   parameter space $\mathbb R_+^p$ with boundaries, which makes    gradient-based influences improper at any boundary point; Example \ref{exa:(Lasso-depth)}
 has a nonsmooth regularizer commonly seen in high dimensional statistics, and sometimes  regularization can be imposed  in a constrained manner.

Following \cite{Rousseeuw1999regression}, the first step   to define a  data depth  is to    characterize a reasonable ``fit'', or a   class of reasonable estimators, under a given model or method. It turns out that for   such nonsmooth problems,
we can    derive  local optimality conditions in   form of \textit{inequalities}  or obtain some \textit{nonlinear} fixed-point equations by use of a surrogate function, neither of which however results in sample-additive estimating equations directly. The good news is that we can then utilize some  ``slack variables''  subject to proper ({convex)} inequality and equality constraints to offer a universal solution, which leads to  a novel class of slacked data depth.

\subsection{Slacked data depth and sparse learning}
\label{subsec:slackspar}
To begin with, let us consider
  $\min f(\boldsymbol{\beta})\triangleq\sum_{i}l(\boldsymbol{\beta};\boldsymbol{x}_{i},y_{i}) \mbox{ s.t. } \boldsymbol{\beta}\succeq\boldsymbol{0}$ or  $\bsbb\in\mathbb R_+^p$,   where $l$ is differentiable  in the augmented parameter space $\mathbb R^p$  but not necessarily convex.  
Because $f$ is   directionally differentiable in $\mathbb R_+^p$,   any optimal solution $\hat \bsbb$ must obey
\begin{align*}
\mathrm{D}_{\boldsymbol{u}}f(\hat\bsbb)\geq0 \mbox{ for all feasible }  \boldsymbol{u}
\end{align*}
where  $\mathrm{D}_{\boldsymbol{u}}f_{}(\boldsymbol{\beta})$
denotes the \textit{one-sided} directional derivative of $f_{}$ at $\boldsymbol{\beta}$
with increment $\boldsymbol{u}$, namely,  $\mathrm{D}_{\boldsymbol{u}}f_{}(\boldsymbol{\beta})=\lim_{\epsilon\rightarrow0+}[f_{}(\boldsymbol{\beta}+\epsilon\boldsymbol{u})-f(\boldsymbol{\beta})]/\epsilon$. Nevertheless, unlike equalities that are  maintained after    projection   (i.e., $\sum {\bsb{T}}_i(\bsbB) = \bsb0\Rightarrow \sum \langle \bsbV, {\bsb{T}}_i(\bsbB)\rangle = \langle \bsbV,\sum  {\bsb{T}}_i(\bsbB)\rangle = 0, \forall \bsbV\in \mathcal G$),  applying the same operation on   inequalities may destroy their meanings totally during  the process of Tukeyfication. 

Our proposal  is   to associate each inequality with an additional slack variable, and append   a nonnegative constraint when performing projection and error measurement.
Let $\boldsymbol{e}_{j}$ be a vector with the $j$th component   $ 1$ and the   remaining   0. In Example \ref{exa:(Non-negative-regression-data)},
  taking $\boldsymbol{u}= \pm\boldsymbol{  e}_{j}$  for $j\in\mathcal{J}= \{j: \beta_j \ne 0\}$  and  $\boldsymbol{u}=\boldsymbol{e}_{j}$
for $j\in{\mathcal{J}}^{c}$ leads to  the following slacked estimating equation: $$\sum_{i}(\nabla l_{}(\boldsymbol{\beta};\boldsymbol{x}_{i},y_{i})-\boldsymbol{s}/n)=\boldsymbol{0},$$ where  $\boldsymbol{s}_{\mathcal{J}^{c}}\succeq\boldsymbol{0}$
and $\boldsymbol{s}_{\mathcal{J}}=\boldsymbol{0}$.
The ordinary   Tukeyfication now goes through, and we obtain a depth optimization problem for any   $\bsbb^\circ\succeq \bsb0$:   $$\min_{(\boldsymbol{v},\boldsymbol{s})\in\mathbb{R}^{p}\times\mathbb{R}^{p}}\sum_{i}1_{\ge 0}(\langle\boldsymbol{v}, \allowbreak\nabla l(\boldsymbol{\beta}^{\circ};\boldsymbol{x}_{i},y_{i})-\boldsymbol{s}/n\rangle)\;\textrm{s.t.}\;\lVert\boldsymbol{v}\rVert_{2}=1,\;\boldsymbol{s}\circ\boldsymbol{\beta}^{\circ}=\boldsymbol{0},\;\boldsymbol{s}\succeq\boldsymbol{0}.$$  When $l(\boldsymbol{\beta};\boldsymbol{x}_{i},y_{i})=(\boldsymbol{x}_{i}^{T}\boldsymbol{\beta}-y_{i})^{2}/2$,
we get the nonnegative regression depth
\begin{equation}
\begin{gathered}\min_{(\boldsymbol{v},\boldsymbol{s})\in\mathbb{R}^{p}\times\mathbb{R}^{p}}\sum_{i}1_{\ge 0}(\langle\boldsymbol{v},\boldsymbol{x}_{i}( \boldsymbol{x}_{i}^{T}\boldsymbol{\beta}^{\circ} - y_{i})-\boldsymbol{s}/n\rangle) \textrm{ s.t. }\lVert\boldsymbol{v}\rVert_{2}=1,\boldsymbol{s}\circ\boldsymbol{\beta}^{\circ}=\boldsymbol{0},\boldsymbol{s}\succeq\boldsymbol{0}.
\end{gathered}
\label{eq:PHDS-non-negative}
\end{equation}
Recall that  $\circ $ denotes the elementwise product.
 When $\boldsymbol{\beta}^{\circ}\succ\boldsymbol{0}$,  $\boldsymbol{s}=\boldsymbol{0}$,
and   (\ref{eq:PHDS-non-negative}) becomes  the  regression depth. In general, the inclusion of $\bsb{s}$  in the minimization, as an outcome of the  nonnegativity restriction, often results in a lower depth value.
\\

The slack-variable technique can introduce useful depth notions for
    sparse learning that is at the core of high dimensional statistics: \begin{equation}
\min_{\boldsymbol{\beta}} f(\bsbb) \triangleq \bar l  ( \bsbX \boldsymbol{\beta}; \bsby)+\sum_{j=1}^{p}P(\lvert\beta_{j}\rvert;\lambda), \label{eq:theta estimator optimization problem}
\end{equation}
where      $\bar l(\bsbX\boldsymbol{\beta};   \bsby) = \sum_{i}l_{0}(  \bsbx_i^T \boldsymbol{\beta}; \boldsymbol{y}_{i})$ with  $ l_0$ differentiable. Here, we assume that     $P$  is   {sparsity-promoting} in the sense that it is induced by a   \textit{{thresholding} rule} $\Theta(\cdot; \lambda)$ with $\lambda$ as the threshold  (see \cite{She2012} for the rigorous definition and more   details):
$ 
P(t;\lambda)=P_{\Theta}(t;\lambda)+q(t;\lambda)$, where
$$P_{\Theta}(t;\lambda) =\int_{0}^{\lvert t\rvert}(\Theta^{-1}(u;\lambda)-u)\rd u \mbox{ with }  \Theta^{-1}(u;\lambda)=\sup\{t:\Theta(t;\lambda)\leq u\}$$ 
 and  $q$ is an arbitrary nonnegative function satisfying $q(t;\lambda)=0$ if $t=\Theta(s;\lambda)$ for some $s\in\mathbb{R}$.
Hence if $\Theta(\cdot;\lambda)$ is a continuous function, $q$ must be identical to zero, but if $\Theta$ has discontinuities, the mapping from    $P$ to  $\Theta$ is   \textit{many-to-one}. The universal $\Theta$-$P$ framework covers many practically used   penalties such as $\ell_r$  ($0\le r \le 1$), SCAD, MCP, which can be nonconvex. For centered  response and predictors,  \eqref{eq:theta estimator optimization problem} suffices; when    centering the response is inappropriate,  an intercept $\alpha$ subject to no regularization should often be added in the systematic component. For clarity, we assume $\alpha =0$ in the following derivation, but the extension to $\bsbX \bsbb + \alpha \bsb1$ is straightforward.

For penalties with  $q\equiv 0$ (continuous $\Theta$), like $\ell_1$ and  SCAD, we can  use
the directional derivatives   along      $\pm \bsb{e}_j$   to show that any locally optimal   $\hat \bsbb$ satisfies the thresholding equation  \citep{she2016c}
\begin{equation}
 {\boldsymbol{\beta}}= {\Theta}( {\boldsymbol{\beta}}-  \bsbX^T \nabla {\bar l}(\bsbX  { \boldsymbol{\beta}});\lambda) ,\label{eq:Theta-eq}
\end{equation}
under the mild assumption that  $\Theta( \cdot ;\lambda)$ is continuous  at  $\hat {\boldsymbol{\beta}}-  \bsbX^T \nabla {\bar l}(\bsbX \hat { \boldsymbol{\beta}})$. But   nontrivial $q$'s and discontinuous $\Theta$'s  constitute  an important   class of nonsmooth  penalties, including, in particular, the discontinuous $\ell_0$ penalty $$\frac{\lambda^2}{2} \| \bsbb\|_0,$$ for which $\Theta$ is the hard-thresholding $\Theta_H(t;\lambda) = t 1_{|t|> \lambda}$, and $q(t;\lambda)= (1/2) (\lambda-|t|)^2    1_{0 < |t| <
\lambda} $. In such scenarios,   if $\nabla \bar l$ is  $L$-Lipschitz continuous, the solutions can be characterized by the  \emph{fixed points} of an iterative optimization algorithm based on a surrogate function $g$:
$$
\beta \in \arg\min  g(\cdot, \bsbb^-)|_{\bsbb^-=\bsbb}
$$
where  $g(\bsbb, \bsbb^-) =\bar l (\bsbX\bsbb^-) + \langle \nabla \bar l (\bsbX\bsbb^-) , \bsbX(\bsbb - \bsbb^-)\rangle+\sum P(\lvert\beta_{j}\rvert;\lambda)  + \rho \| \bsbb - \bsbb^-\|_2^2/2  $ with  $\rho\ge L\| \bsbX\|_2^2$, and we can prove that they all  enjoy nearly minimax error rate under a proper choice of $\lambda$ and  some regularity conditions \citep{SheetalBreg}. 
 It can be verified that      the fixed-point     estimators satisfy \eqref{eq:Theta-eq} as well, under the same continuity assumption, as long as $\bsbX$ has been properly scaled:   $\| \bsbX\|_2 \le  1/\sqrt L$. 
 Unfortunately, the  thresholding equation
  does not belong to the estimating equation framework  examined in our companion paper---specifically,  the nonlinear thresholding effect desired in sparse learning  means that  \eqref{eq:Theta-eq} is not sample additive.

  Below we introduce  $p$  additional slack variables to find a proper substitute for    \eqref{eq:Theta-eq}   so that one can define  data depth for \eqref{eq:theta estimator optimization problem} given  an {\textit{arbitrary}} thresholding
  $\Theta$. Let  $\bsbb$ be a locally optimal solution to the problem as $P=P_\Theta$, or a fixed-point solution as $P=P_\Theta+q$. Define  $\mathcal{J}=\{j:\beta_{j}\neq0\}$
and $\mathcal{J}^{c}=\{j:\beta_{j}=0\}$, and  denote by     $\bsbX[,j]$  the $j$th column vector of $\bsbX$. Using  the directional derivatives of $f$ when  $P=P_\Theta$   (see, e.g., the proof of Theorem 1 in \cite{she2016c}), or  the directional derivatives of $g-q$   under the continuity assumption when $P=P_\Theta+q$,  we get
\begin{align}{\Theta}^{-1}(|{\beta}_{j}|;\lambda) \sgn({\beta}_{j})= {\beta}_{j}- \bsbX[,j]^T\nabla {\bar l}(\bsbX \boldsymbol{\beta}), \  \forall j\in\mathcal{J}, \label{thetaeq-nz}
\end{align}
  which holds even if $\Theta$ is not strictly increasing in a neighborhood of $|\beta_j|$ ($j\in \mathcal J$), while for $j\in\mathcal{J}^{c}$, $\beta_{j}=0$, and so \begin{align}-\lambda \le   \bsbX[,j]^T \nabla {\bar l}(\bsbX \boldsymbol{\beta}) \leq\lambda, \forall j\in\mathcal{J}^{c}.\label{thetaeq-zz}\end{align} 
 Next, define    $\bsbg(\bsbb)=[\gamma_j]$ with
\begin{align}\gamma_{j}=  \begin{cases}\Theta^{-1}(|\beta_{j}|;\lambda)\sgn(\beta_{j})-\beta_{j} & \mbox{ if } j\in\mathcal{J}\\
   0 & \mbox{ if } j\in\mathcal{J}^c.\end{cases} \end{align}         It follows from \eqref{thetaeq-nz}, \eqref{thetaeq-zz} that     $ \bsbX^T \nabla {\bar l}(\bsbX \boldsymbol{\beta})+\boldsymbol{\gamma}(\bsbb)+\boldsymbol{s}=\boldsymbol{0}$
 for some  $\boldsymbol{s}\in\mathbb{R}^{p}$, $\boldsymbol{s}_{\mathcal{J}}=\boldsymbol{0}$, and $|s_j|\leq\lambda$, $j \in \mathcal J^c$.

Now, given a penalty induced by a thresholding rule   $\Theta(\cdot; \lambda)$ and a point of interest $\boldsymbol{\beta}^{\circ}\in\mathbb{R}^{p}$,
the  slacked data depth resulting from  \eqref{eq:theta estimator optimization problem}, which we call  ``$\Theta$-depth'',     can be cast as a \textbf{joint} optimization problem with respect to    direction $\boldsymbol{v}\in \mathbb R^p$
and   slack variables $ \bsb{s}= [{s}_j]\in \mathbb R^p$:
\begin{equation}
\begin{split}
\mbox{\textbf{$\Theta$-depth:}} \quad d_{  01}^{\Theta}(\boldsymbol{\beta}^{\circ})=   \min_{(\boldsymbol{v},\boldsymbol{s})} & \sum_{i}1_{\ge0}(\langle\boldsymbol{v}, \bsbx_i   l_0'(\bsbx_i^T \boldsymbol{\beta}^{\circ}; y_{i})+(\boldsymbol{\gamma}^{\circ}+\boldsymbol{s})/n\rangle)   \\ & \textrm{ s.t. } \lVert\boldsymbol{v}\rVert_{2}=1,\;\boldsymbol{s}\circ\boldsymbol{\beta}^{\circ}=\boldsymbol{0},\;\lVert\boldsymbol{s}\rVert_{\infty}\leq\lambda,
\end{split}
\label{eq:PHDS}
\end{equation}
where      $\bsbg^\circ=\bsbg(\bsbb^\circ)$ and $\| \bsb{s}\|_{\infty} = \max |s_j|$.
When $\Theta$ is the hard thresholding $\Theta_H$ (corresponding to the class of  $\ell_0$ penalties), $\bsbg^\circ = \bsb0$. The user should specify a reasonably small  $\lambda$ (otherwise extremely low depth  values are to be expected):   a theoretical choice in sparse regression is $\lambda = \sigma \sqrt{c n \log p}$ (with say $c=2$) where $\sigma$ is the Orlicz $\psi_2$-norm of the noise, and a less conservative one can often be obtained via cross-validation.
A fascinating fact is that   the slack variable approach requires no  convexity of either the loss  or the penalty.

  An important alternative to  penalized sparse learning is to directly limit the sparsity level:   $\| \bsbb\|_0\le  q$,  instead of specifying a penalty parameter $\lambda$.
Due to the lack of  nonsmoothness of
\begin{align}\min_{ \| \bsbb\|_0\le q}  \bar l  ( \bsbX \boldsymbol{\beta}; \bsby) , \label{l0constrprob}
\end{align} we take the surrogate route. Statistically accurate estimates can be obtained from     the resulting iterative quantile-thresholding   algorithm \citep{SheetalPIQ},  which all satisfy   $ {\boldsymbol{\beta}}= {\Theta}^\# ( {\boldsymbol{\beta}}- (1/  \rho) \bsbX^T \nabla {\bar l}(\bsbX  { \boldsymbol{\beta}}) ;q), $ assuming no ties occur and  $\rho$ is large enough (e.g.,  $ L\| \bsbX\|_2^2  $). Here,  the quantile thresholding    $\Theta^{\#}(\bsb{\alpha}; q )$  for any   $\bsb{\alpha} \in \mathbb R^p$ is  a vector $\bsb{\zeta}$ with
$\zeta_{( j )} = \alpha_{( j )}  \mbox{ if } 1 \leq j \leq q \mbox{, and } 0 \mbox{ otherwise,}$
where $\alpha_{(1)}, \ldots, \alpha_{(p)}$ are the order statistics of $\alpha_1, \ldots, \alpha_p$ satisfying $|\alpha_{(1)}| \geq \cdots \geq |\alpha_{(p)}|$. $\Theta^\#(\bsbalpha; q)$ can be viewed as   a variant of  $\Theta_H(\bsbalpha; \lambda)$ (by setting $\lambda =|\alpha_{(q+1)}|$, say), but it uses an adaptive threshold. Again, we suppose that the regularization parameter  $q$ is already given.

By use of slack variables to rewrite the $\Theta^\#$-equation (details omitted), we can define the $q$-sparse constrained $\ell_0$-depth (which we call  ``$\Theta^\#$-depth'') for any   $\bsbb^\circ: \|\bsbb^\circ\|_0= q$  as \begin{align}
\begin{split}
\mbox{\textbf{$\Theta^\#$-depth:}} &\  \min_{(\boldsymbol{v},\boldsymbol{s})\in\mathbb{R}^{p}\times\mathbb{R}^{p}}   \sum_{i}1_{\ge0}(\langle\boldsymbol{v}, \bsbx_i   l_0'(\bsbx_i^T \boldsymbol{\beta}^{\circ}; y_{i})+ \boldsymbol{s} /n\rangle)  \\ & \mbox{ s.t. } \lVert\boldsymbol{v}\rVert_{2}=1,\;\boldsymbol{s}\circ\boldsymbol{\beta}^{\circ}=\boldsymbol{0},\;\lVert\boldsymbol{s}\rVert_{\infty}\le \|   \bsbX_{({\mathcal J}^{\circ  })^c}^T \nabla {\bar l}(\bsbX  { \boldsymbol{\beta}}^\circ)\|_{\infty},    
    \end{split}\label{sparsitydepthdef}
\end{align}
where $({\mathcal J}^{\circ })^c = \{j: \beta_j^\circ = 0\}$,   $ \bsbX_{({\mathcal J}^\circ)^c}$ is a submatrix of $\bsbX$ by selecting the columns corresponding  to the complement of ${\mathcal J}^\circ$,  
and both  $\bsbg^\circ$ and      $\rho$ (as long as $\rho>0$)   disappear in the   $\ell_0$-constrained  depth, just  like in the   $\ell_0$-penalized case.  A similar derivation is presented in detail in Section \ref{subsec:rrrdepth}. Of all  the constraints on $\bsb{s}$,  the equality ones are affine, and the inequality ones are convex. The deepest $q$-sparse estimate is defined as the saddle point that maximizes \eqref{sparsitydepthdef} over all $\bsbb^\circ: \|\bsbb^\circ\|_0= q$ (cf. \eqref{eqcompositedep} in \cite{SheDepthI2021}).

Clearly, in the special case of  $q=p$,    all slack variables are removed, but as $q<p$, the constrained problem \eqref{l0constrprob} results in  more stringent estimating equations that are easier to violate, compared with the plain (non-regularized) problem. This is  reflected by the inclusion of $\bsb{s}$ during the minimization, thereby  lower depth values.  The same conclusion holds   for the  $\Theta$-depth \eqref{eq:PHDS} due to the existence of additional  slack variables. On the other hand,  sparsity depths may be very \textbf{low}  for large   $p$  and    $\lambda$  in  \eqref{eq:PHDS} or  large $p$ and small $q$  in \eqref{sparsitydepthdef}. To alleviate the issue, it is beneficial to change  the crude   ``0-1 loss'' to some more elegant $\varphi$, as discussed in our companion paper \citep{SheDepthI2021}. For example,     \eqref{sparsitydepthdef} could be replaced by
$ \min_{(\boldsymbol{v},\boldsymbol{s})}   \sum_{i} \varphi (\langle\boldsymbol{v}, \bsbx_i   l_0'(\bsbx_i^T \boldsymbol{\beta}^{\circ}; y_{i})/\rho+ \boldsymbol{s} /n\rangle)   \mbox{ s.t. } \lVert\boldsymbol{v}\rVert_{2}=1,\;\boldsymbol{s}\circ\boldsymbol{\beta}^{\circ}=\boldsymbol{0},\;\lVert\boldsymbol{s}\rVert_{\infty}\le \|   \bsbX_{({\mathcal J}^{\circ  })^c}^T \nabla {\bar l}(\bsbX  { \boldsymbol{\beta}}^\circ)/\rho\|_{\infty} $, with    $\varphi(\cdot)$  nonzero for mild or moderate  negative inputs, which warrants further investigation in the future. A tight upper bound of sparsity depths is also worth studying in theory.

Slacked data depth can be introduced for groupwise variable selection and low-rank matrix estimation \citep{She2012,SheTISPMat} as well; see, e.g., Section \ref{subsec:rrrdepth}. 

\begin{remark}[Computation of slacked depth] \label{rmk:compslackdepth}
A simple alternating optimization or   \emph{block coordinate descent} (BCD) algorithm can be used to to compute slacked data depth. Take  the $\varphi$-form of (\ref{eq:PHDS}) as an example. Given $\boldsymbol{s}$, the optimization
problem for $\boldsymbol{v}$, $$\min_{\boldsymbol{v}\in\mathbb{R}^{p},\lVert\boldsymbol{v}\rVert_{2}=1}\allowbreak\sum_{i}\varphi (\langle\boldsymbol{v},r_{i}\boldsymbol{x}_{i}+(\boldsymbol{\gamma}^{\circ}+\boldsymbol{s})/n\rangle) $$ where $\boldsymbol{r}=\nabla_{\boldsymbol{\Theta}}\bar{l}\lvert_{\boldsymbol{\Theta}=\boldsymbol{X}\boldsymbol{\beta}^{\circ}}$,
  has been  investigated  in \cite{SheDepthI2021}.
Fixing $\boldsymbol{v}$, we can rewrite the $\bsb{s}$-problem as
\[
\min_{\boldsymbol{s}\in\mathbb{R}^{p}}\sum_{i}\varphi (\langle\boldsymbol{v},r_{i}\boldsymbol{x}_{i}+\boldsymbol{\gamma}^{\circ}/n+\boldsymbol{s}/n\rangle)\; \textrm{  s.t. }\; \bsb{s}_{\mathcal J} = \bsb{0},\lVert\boldsymbol{s}\rVert_{\infty}\leq\lambda.
\]
The problem has a differentiable criterion in $\boldsymbol{s}$ and some simple
box constraints, and     conventional numerical methods apply, including L-BFGS-B,    interior point, and proximal gradient descent algorithms \citep{Byrd1995,Boyd2004,Parikh2014}.
\end{remark}

\subsection{Reduced-rank regression depth}
\label{subsec:rrrdepth}
Applying ordinary least squares  on multiple responses may easily result in a large number of unknowns. Researchers  often prefer adding a low-rank constraint in estimating the coefficient matrix, leading to the celebrated reduced-rank regression (RRR) \citep{And51}
\begin{align}\min_{\bsbB\in \mathbb R^{p\times m}}f(\bsbB; \bsbX, \bsbY)\triangleq \frac{1}{2}\|\bsbY - \bsbX \bsbB\|_F^2  \mbox{ s.t. } \mbox{rank}(\bsbB)\leq r, \label{rrr}\end{align} where   $\bsbY=[\boldsymbol{y}_{1}\ldots\boldsymbol{y}_{n}]^{T}\in \mathbb R^{n\times m}$ and $\bsbX=[\boldsymbol{x}_{1}\ldots\boldsymbol{x}_{n}]^{T}\in \mathbb R^{n\times p}$ are the (centered)   response and predictor matrices.   A weighted criterion to account for the dependency between the responses can be given, but the problem can be converted to  \eqref{rrr} with a simple reparametrization. If   the variables are not centered, an intercept term $\bsb{1} \bsb{\alpha}^T$  should  be added in the loss, but the depth derivation below carries over (cf. Section \ref{subsec:depthpca}). We assume that $(\bsbx_i, \bsby_i)$ are   i.i.d. (or in an approximate sense),  and so the data depth in this subsection does not apply to PCA  where $\bsbX = \bsbI$, thus distinct from the   PC-depth and OC-depth introduced earlier; see some related discussions in  Section \ref{subsec:ex} of  our companion paper.

RRR   provides a  low-dimensional projection  space to view and analyze supervised multivariate  data, and finds widespread applications in  machine learning and   econometrics  \citep{ReinVelu,izenbook}. In fact, once an estimate $ \bsbB$ of rank $r$ is obtained, we can   write    $ \bsbB = \bsbB_1 \bsbB_2^T$ for   $\bsbB_1\in \mathbb R^{p\times r}$, $\bsbB_2\in \mathbb R^{m\times r}$. This suggests that    $r$  factors    can be  constructed  by  $\bsbX\bsbB_1$ from   $p$  predictors  to  explain all  response variables. The number of  factors required in real applications is often much smaller than the number of input $x$-variables.

Limiting the rank of  the matrix estimators at $r$, how to perform a ``center-outward'' ranking  in high dimensions, or more generally, test $$H_0: \bsbB\in \Omega_0 \cap \{ \mbox{rank}(\bsbB) =r\} \mbox{ vs. }H_a: \bsbB  \in \Omega_0^c \cap \{ \mbox{rank}(\bsbB) =r\},$$ where the set or event $\Omega_0$ is not necessarily a singleton (cf. Remark \ref{rem:depth median computation} in \cite{SheDepthI2021}), is an intriguing  open question.

In the following, we   extend  multivariate regression depth \citep{Rousseeuw1999regression,Bern2002}  to the \textit{reduced-rank regression depth}   \eqref{rrrdepth}, using the techniques developed in the last subsection. 
Toward this,  we first   give a fixed-point formulation of all RRR estimators.  Define a matrix version of the $\Theta^\#$  introduced in the last subsection
\begin{equation}
\Theta^{\sigma\#} (\boldsymbol{B};r) \triangleq \bsbU{\rm{diag}}\{ \Theta^{\#}([\sigma_i{(\bsbB)}] ; r) \}  \bsbV^T, \ \ \ \forall \bsbB \in \mathbb{R}^{p \times m}\label{eq:matqth}
\end{equation}
where $\bsbU$, $\bsbV$, and diag$\{\sigma{(\bsbB)}_i\}$ are  from  the SVD of $\bsbB = \bsbU \text{diag}\{ \sigma_i{(\bsbB)} \} \bsbV$, and  $\Theta^\#$  is applied to the vector $[\sigma_i{(\bsbB)}]$, with $\sigma_i{(\bsbB)}$ denoting the $i$th largest singular value of $\bsbB$.

Construct a surrogate function  $$g(\bsbB, \bsbB^-) =f(\bsbB^-) + \langle \nabla f(\bsbB^-), \bsbB - \bsbB^-\rangle  + \rho \| \bsbB - \bsbB^-\|_F^2/2,  $$ where   $\rho$ is  larger than $\| \bsbX\|_2^2$.  Let  $\hat \bsbB_{\mbox{\tiny rrr}}$ be an RRR estimator that solves
\eqref{rrr}. Then for $\tilde \bsbB \in \arg\min_{\bsbB: \mbox{\tiny rank}(\bsbB)\le r} g(\bsbB, \hat \bsbB_{\mbox{\tiny rrr}})$, it follows from  the chain inequalities $f(\hat\bsbB_{\mbox{\tiny rrr}}) - f(\tilde \bsbB)= g(\hat \bsbB_{\mbox{\tiny rrr}}, \hat \bsbB_{\mbox{\tiny rrr}})- f(\tilde \bsbB)\ge g(\tilde \bsbB, \hat \bsbB_{\mbox{\tiny rrr}}) - f(\tilde \bsbB)\ge (\rho - \| \bsbX\|_2^2) \| \tilde \bsbB - \hat \bsbB_{\mbox{\tiny rrr}}\|_F^2/2$ that  $\hat \bsbB_{\mbox{\tiny rrr}} = \tilde \bsbB$. On the other hand, it is easy to show that $\tilde \bsbB= \Theta^{\sigma\#} ( \hat \bsbB  _{\mbox{\tiny rrr}}- \frac{1}{\rho} \boldsymbol{X}^T(\bsbX\hat \bsbB_{\mbox{\tiny rrr}}-\bsb{Y});r) $ \citep{SheTISPMat}, and so $\hat \bsbB_{\mbox{\tiny rrr}}$ satisfies the matrix thresholding equation
\begin{equation}  \bsbB  = \Theta^{\sigma\#}(  {\bsbB} - \frac{1}{ \rho}\boldsymbol{X}^T(\bsbX \bsbB-\boldsymbol{Y});r). \label{rrr-thetaeq}
\end{equation}
(In fact, under the mild condition that $\bsbY^T  \bsbX (\bsbX^T\bsbX)^+\bsbX^T\bsbY$     has distinct eigenvalues, the RRR estimator is unique \citep{ReinVelu}, and $\rho$ can be way smaller than  $\| \bsbX\|_2^2$.) 
Perform a compact SVD of  $\bsbB$:  $\bsbB = \boldsymbol{P}  \boldsymbol{D} \boldsymbol{Q}^T$   with $\bsbP\in \mathbb O^{p\times r}$ and $\bsbQ \in \mathbb O^{m\times r}$, and denote by  $\bsbP_\perp\in \mathbb O^{p\times (p-r)}$ and $\bsbQ_\perp\in \mathbb O^{m\times (m-r)}$  their orthogonal complements (which can  be  obtained from the full SVD of $\bsbB$).  Like in  the $\Theta^\#$-case, based on
\eqref{rrr-thetaeq}, we  work on  $   ({1}/{ \rho})\boldsymbol{X}^T(\bsbX   \bsbB -\boldsymbol{Y})   + \bsbS=\bsb0 $ for a slack matrix $\bsbS$ satisfying $$\boldsymbol{P}^T\bsbS=\boldsymbol{0} ,  \bsbS \boldsymbol{Q} = \boldsymbol{0} ,  \| \bsbS\|_2  \le \| \frac{1}{\rho}\bsbP_{\perp}\bsbP_{\perp}^T\boldsymbol{X}^T(\bsbX   \bsbB -\boldsymbol{Y}) \bsbQ_{\perp} \bsbQ_{\perp}^T\|_{2}.$$

Now, given  a regularization parameter $r: 1\le r\le p\wedge m$ and  a   matrix of interest $\bsbB^\circ\in \mathbb R^{p\times m}: \mbox{rank}(\bsbB^\circ)= r$,   obtain the associated    $\bsbP_\perp^\circ\in \mathbb O^{p\times (p-r)}$, $\bsbQ_\perp^\circ\in \mathbb O^{m\times (m-r)}$ as above;
the rank-$r$  RRR depth of $\bsbB^\circ$  is   defined by
\begin{align*}
&d_{01}^{\mbox{{\fontsize{4.5}{2}\selectfont RRR}}}(\bsbB^\circ) =  \min_{(\bsbV,\bsbS)} \sum_i 1_{\geq 0} (\langle \bsbV,   \frac{1}{\rho}\bsbx_i(\bsbx^T_i\bsbB^\circ - \bsby_i^T) + \frac{1}{n} \bsbS \rangle ) \\ & \ \text{ s.t. } \| \bsbV \|_F = 1, \boldsymbol{P}^{\circ T}\bsbS=\boldsymbol{0}, \bsbS\boldsymbol{Q}^{\circ}=\boldsymbol{0}, \|\bsbS \|_2 \leq  \| \frac{1}{\rho}\bsbP_{\perp}^{\circ T} \boldsymbol{X}^T(\bsbX   \bsbB^\circ-\boldsymbol{Y}) \bsbQ_{\perp}^{\circ} \|_{2},
\end{align*}
or equivalently,
\begin{align}
\mbox{\textbf{RRR-depth:}}\ \ &  d_{01}^{\mbox{{\fontsize{4.5}{2}\selectfont RRR}}}(\bsbB^\circ) =  \min_{(\bsbV,\boldsymbol{L})} \sum_i 1_{\geq 0} (\langle \bsbV,    \bsbx_i(\bsbx^T_i\bsbB^\circ - \bsby_i^T) + \frac{1}{n} \boldsymbol{P}_\perp^{\circ}\boldsymbol{L}\boldsymbol{Q}^{\circ T}_\perp \rangle  ) \notag \\ & \  \text{  s.t. } \| \bsbV \|_F = 1, \|\boldsymbol{L} \|_2 \leq   \| \bsbP_{\perp}^{\circ T} \boldsymbol{X}^T(\bsbX   \bsbB^\circ-\boldsymbol{Y}) \bsbQ_{\perp}^{\circ} \|_{2},
 \label{rrrdepth}
\end{align}
where $\rho$ vanishes due to the scale invariance of $1_{\ge 0}$, regardless of how small or large $\rho$ is. Clearly, in the full rank case $r = m\wedge p$, either $\bsbP_\perp^{\circ}$ or $\bsbQ_\perp^{\circ}$  must vanish, and so  $\bsbL = \bsb0$, meaning that   \eqref{rrrdepth} reduces to the multivariate regression depth \citep{Bern2002}.

\begin{remark}[Combined treatment]
The manifold   approach and   slack variable approach can be combined together to define data depth for some challenging problems. Consider a \emph{sparse} RRR (one of the    variants in \cite{She2017}) that  constructs  $r$ predictive factors from  a  subset of  predictors
\begin{align}
\min_{\bsbS\in \mathbb R^{p \times r}, \bsbU\in \mathbb R^{m\times r}} \frac{1}{2}\|\bsbY - \bsbX \bsbA \bsbU^T\|_F^2 \mbox{ s.t. } \|\vect(\bsbA)\|_0 \le q, \  \bsbU ^T \bsbU = \bsbI_{r\times r}. \label{srrr}
\end{align}
The overall coefficient matrix $\bsbB = \bsbA \bsbU^T$ has rank at most $r$ as in RRR, but sparsity is imposed on the loading matrix $  \bsbA$. 
  By use of   a slack   matrix $\bsbS$ for $\bsbA$, and a Riemannian  tangent space for $\bsbU$,  the depth  for $(\bsbA^\circ, \bsbU^\circ)$: $\bsbA^\circ\in \mathbb R^{p\times r}, \|\vect(\bsbA^\circ)\|_0= q$, $\bsbU^\circ\in\mathbb O^{m\times r}$ is given by
$
 \min_{ \bsbW\in \mathbb R^{m\times r}, \bsbV\in  \mathbb R^{p\times r}, \bsbS\in\mathbb R^{p\times r}   }\allowbreak \sum_{i=1}^n   1_{\ge 0} (- \langle \bsbW,  \bsby_i \bsbx_i^T \bsbA^\circ \rangle
 + \langle \bsbV, \bsbx_i (\bsbx_i^T \bsbA^\circ - \bsby_i^T \bsbU^\circ )+ \bsbS/n\rangle ) $ s.t.  $ \|\bsbW\|_F^2 + \| \bsbV\|_F^2=1,\bsbV^T \bsbU^\circ +  \bsbU^{\circ  T} \bsbV = \bsb0,  \vect(\bsbA^\circ)\circ \vect(\bsbS) = \bsb0,\| \bsbS\|_{\max}\le  \lambda^\circ$, with $\lambda^\circ = \|\vect(\bsbX^T(\bsbX \bsbA^\circ - \bsbY \bsbU^\circ))[{({\mathcal J}^{\circ    })^c}]\|_\infty= \|(\bsbI\otimes \bsbX^T)[{({\mathcal J}^{\circ  })^c}, ] \vect(\bsbX \bsbA^\circ - \bsbY \bsbU^\circ)\|_{\infty}$ and ${\mathcal J}^\circ = \{j:  \vect(\bsbA^\circ) [j] \ne 0, 1\le j\le pr \}$.  
\end{remark}

\section{Experiments}
\label{sec:experiments}
This section performs    real data experiments to illustrate the usefulness of some new  notions of depth.
\subsection{Reduced-rank depth in time series}

We consider the 52 weekly stock log-return data for nine of the ten largest American corporations in 2004  \citep{Rothman2010}, with $\bsby_t\in\mathbb{R}^9$ ($t=1,\ldots,T$) and $T=52$.

For the purpose of  constructing market factors that drive general stock movements, a reduced-rank vector autoregressive (VAR) model can be used, i.e., $\bsby_{t+1} =\bsbB^{*T}\bsby_{t}+\bsb{e}_t$, with $\bsbB^*$ of low rank. By conditioning on the initial state $\bsby_0$ and assuming the normality of $\bsb{e}_t$, the conditional likelihood function leads to a least squares criterion, so the estimation of $\bsbB^*$ can be formulated as a reduced-rank regression problem; see \cite{Ltkepohl2007} for more details. We fit the reduced-rank VAR  with
$r=6$. The optimization algorithm for \eqref{rrrdepth} (implemented based on  Remark \ref{rmk:compslackdepth}) however shows that the objective function  can reach zero for some feasible $(\bsbV,\boldsymbol{L})$. Hence, although the standard RRR approach   is  widely used
 in multivariate times series and econometrics, our  analysis
 revealed  a perhaps surprisingly low data depth   on this financial series dataset.

 We then considered the  Cauchy-based reduced rank regression \citep{zhao2017robust,packageRRRR} (denoted by C-RRR) and
robust reduced rank regression \citep{She2017RRRR} (denoted by R4, with 5\% of data treated as  outliers), as well as a deeper estimate obtained by random sampling (denoted by D-RRR). 
The  rank-6 depth values  of these estimates are
0.02, 0.08 and 0.12, respectively, suggesting   more  reliable fitted models  than the plain RRR from the perspective of data depth.

\begin{figure}[ht!]
\begin{center}
{\includegraphics[width=0.8\columnwidth]{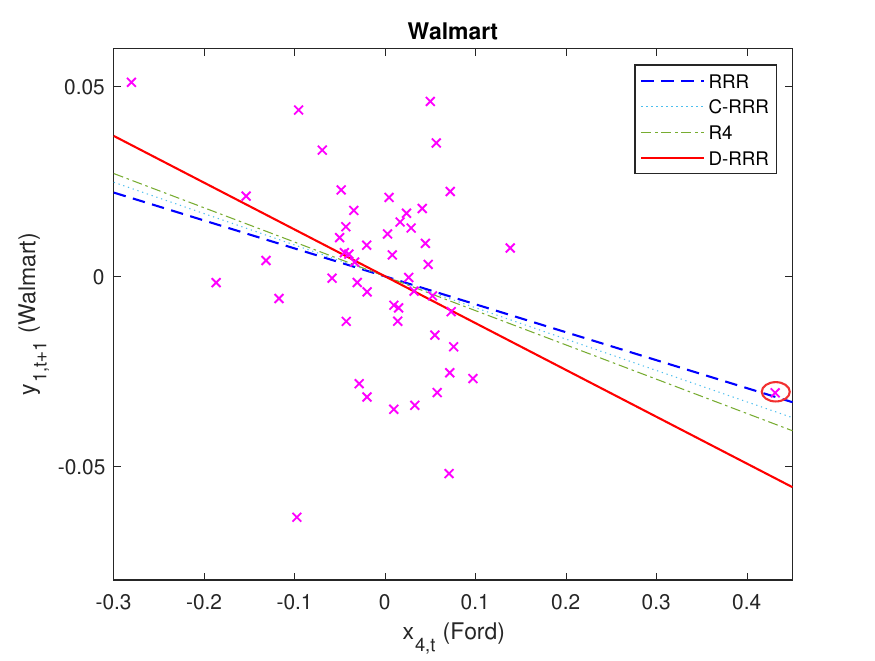}}
\caption{\small The fitted models     $y_{1,t+1}\sim  x_{4,t} $  using different  methods in the low-rank VAR(1), to demonstrate how the  log-return of {\textsc{Walmart}} is related to that of
{\textsc{Ford}} in the previous week. Notice the right-most point that has a high leverage. \label{fig:RRRcoeffs}}
\end{center}
\end{figure}

To further illustrate the differences between the estimates, we plot the fitted models of  {\textsc{Ford}} ($x_{4,t}$) in response to {\textsc{Walmart}}
($y_{1,t+1}$)  in Figure \ref{fig:RRRcoeffs}. Notably,   the right-most point  has  high leverage, and the
RRR  model passes close to that particular observation. In contrast, D-RRR seems to   fit better the majority of the sample.


A careful examination of the  series  shows the point corresponds to  the log-return of \textsc{Ford} at week 17,  a real major market disturbance attributed to the auto industry. Several other stock returns experienced
dramatic
short-term changes as well, and we occasionally observe that the slopes obtained from RRR\ and its robust counterparts     can have opposite signs.
Financial time series
often contain
anomalies or demonstrate heavier tails than those of a normal distribution due to extreme market movements.
The  issue may jeopardize the recovery of common market behaviors and   asset return forecasting:
the autoregressive structure  can make any outlier in the time series also a leverage point in the covariates. Although an elaborate  robustification
of the low-rank VAR
 merits further  investigation,   our depth-based analysis
seems to offer an effective  fix in this regard.

\subsection{Sparsity depth for performance evaluation}
 Data depth provides a nonparametric means of  performance evaluation. In this experiment, we use the sparsity depth defined in \eqref{sparsitydepthdef} to conduct a comparison between some commonly used   sparse learning methods  on the Boston housing dataset    \citep{Harrub1978}.
The dataset was  collected by the U.S. Census Service and consists of  13 predictors regarding socioeconomic and environmental conditions for 506 neighborhoods in the Boston area. The response is the median value of owner-occupied homes in the area.

We   compare  Lasso \citep{Tibshirani1996},          SCAD \citep{Fan2001}, sparse LTS  (S-LTS) \citep{alfons2013sparse},   quantile-SCAD (Q-SCAD) \citep{rqPen19}      and PIQ \citep{SheetalPIQ}, in terms of data centrality defined  in \eqref{sparsitydepthdef} for the same given support size $q$.
More concretely, assuming that the observations are i.i.d., we split the dataset in halves,   fit the methods on the first half,   and then  evaluate their  performance via sparsity depth on the rest half. The whole procedure is repeated 20 times.

\begin{figure}[ht!]
\subfloat[$q=7$]{\includegraphics[width=0.24\columnwidth]{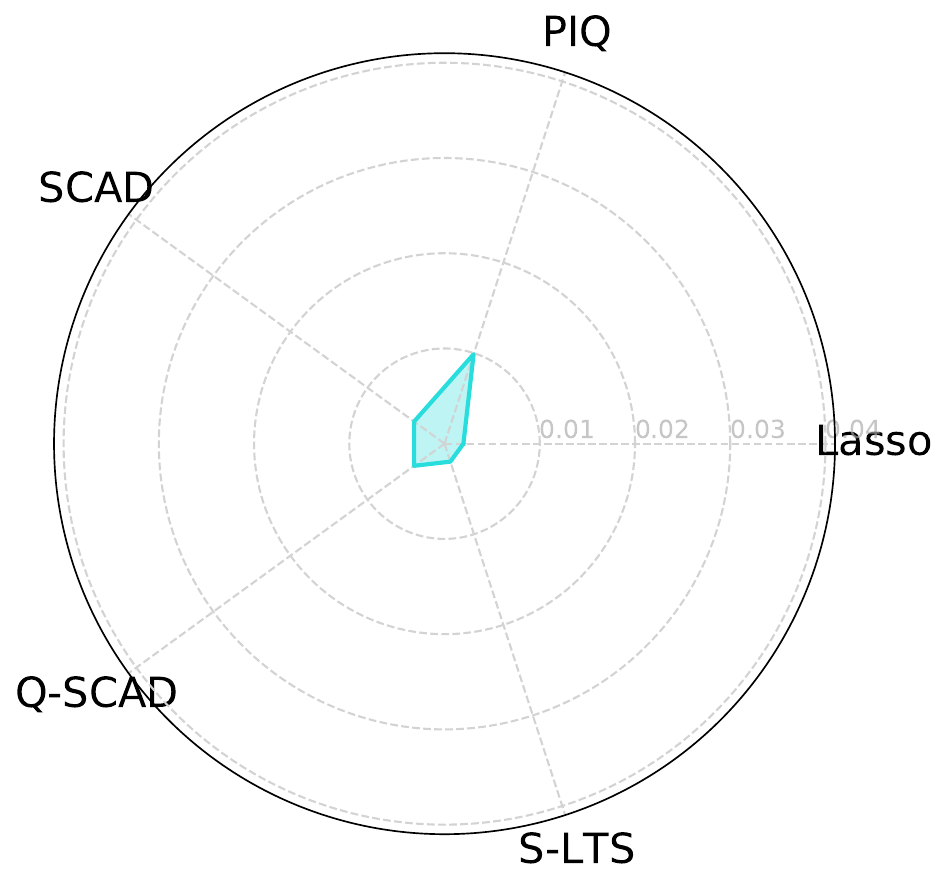}}\
\subfloat[$q=8$]{\includegraphics[width=0.24\columnwidth]{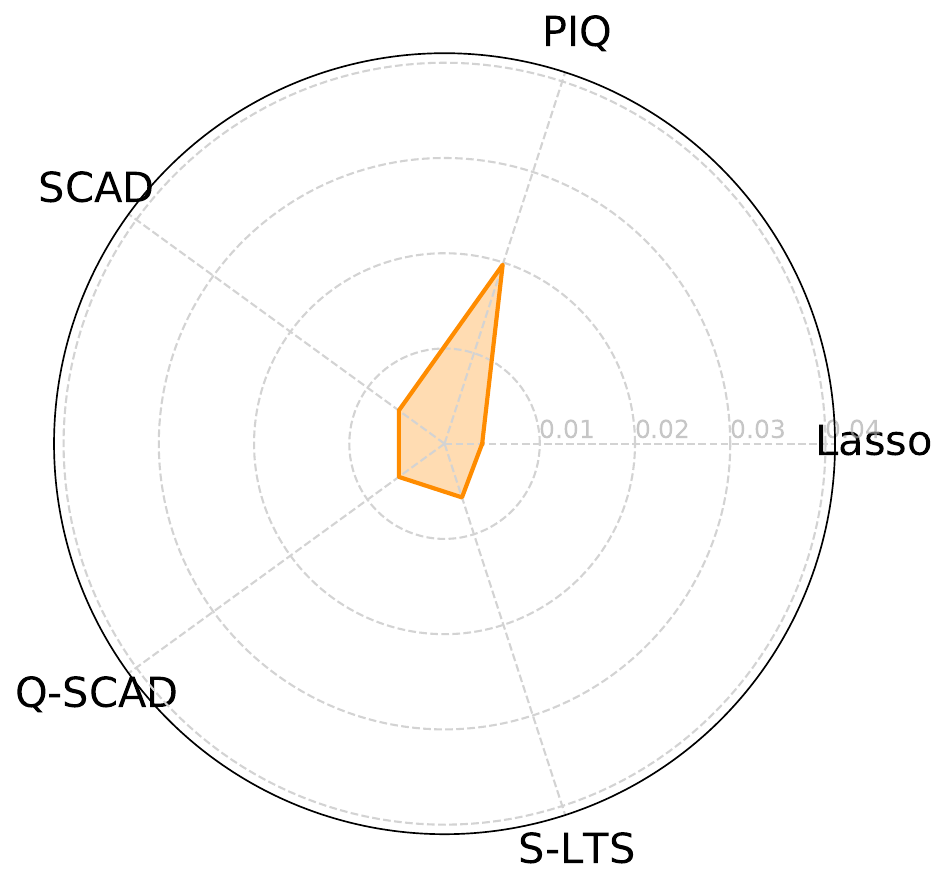}}\
\subfloat[$q=9$]{\includegraphics[width=0.24\columnwidth]{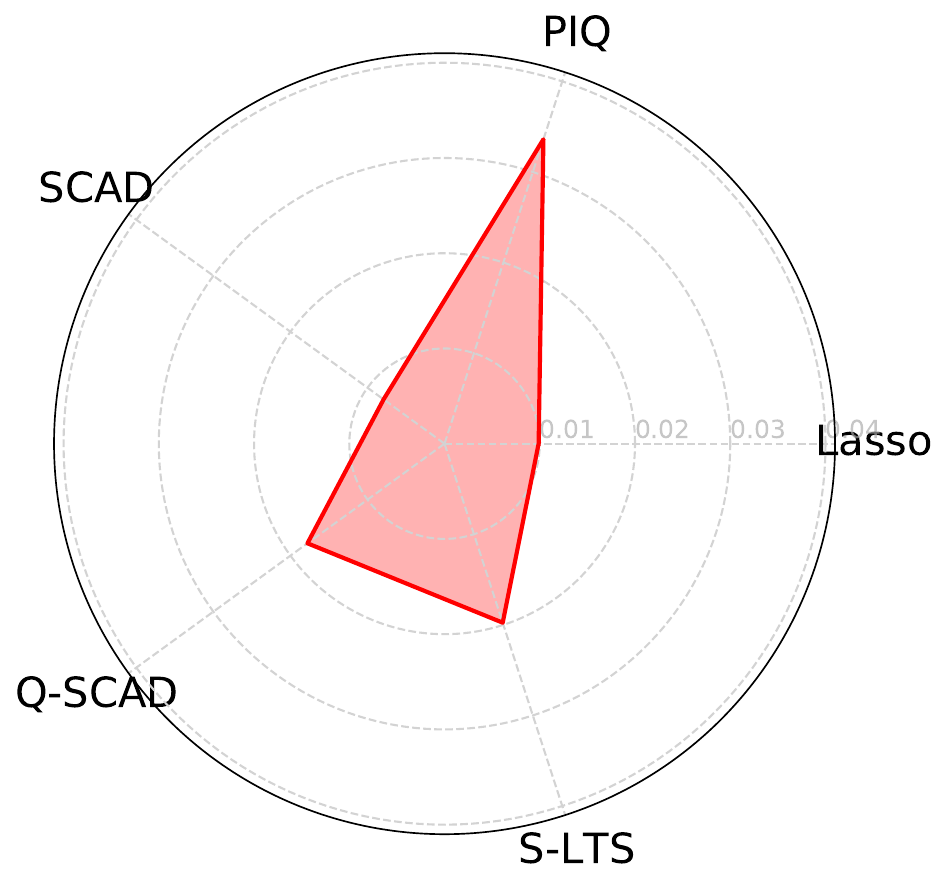}}\
\subfloat[$q=10$]{\includegraphics[width=0.24\columnwidth]{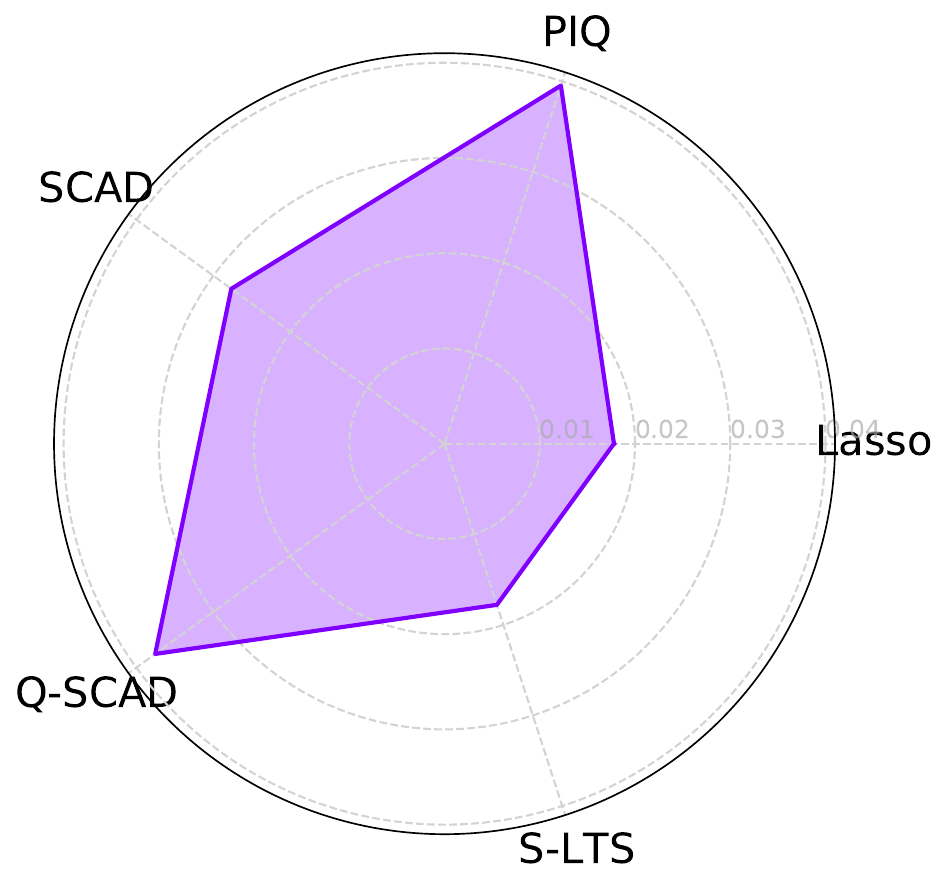}}
\caption{\small Sparsity depth comparison between  Lasso,  SCAD, sparse LTS (S-LTS),    quantile SCAD (Q-SCAD) and PIQ estimates  with respect to the  support size    $q$.  \label{fig:  sparsity depth}}
\end{figure}

 Figure \ref{fig:  sparsity depth} shows a series of radar   plots for the median $\Theta^\#$-depths in respect to the number of selected variables. The depth values are small but present  useful ranking  information. The estimates  are quite      different
seen from the data depth comparison: Lasso and SCAD   exhibit   lower depth in most cases, Q-SCAD
 and PIQ often give  deeper estimates, and S-LTS is      unstable (and costly) in our experiments. The last three methods all  use a more robust loss, as well as
a nonconvex regularizer, while  Lasso solves a convex optimization problem with the ordinary  $\ell_2$-loss and $\ell_1$-penalty. The depth differences between these sparse learning methods       indicate that         the  data  must deviate   from Gaussianity and may contain anomalies, and   incorporating the desired type of regularization  into data  depth  can provide a helpful tool for   robust performance evaluation.

\section{Summary}
\label{sec:summ}
Our work investigated  Tukey's notion of   depth for  robustifying a given optimization criterion, an estimating equation, or an algorithm in  statistical inference and estimation.
In Part I, we introduced a polished subspace depth framework, where the elements like the influence space constraint,    rectified redescending  discrepancy measures,     and subspace projection are new to the best of our knowledge.
In Part II, we   proposed two novel  approaches based on  manifolds and slack variables to   extend the concept    to problems defined in some restricted parameter spaces or with a nonsmooth regularizer.
Our matrix formulation of the problems, together with state-of-the-art optimization techniques (particularly  momentum-based acceleration), gave rise to a new  class of efficient algorithms that has guaranteed  convergence and scales up with  problem dimensions. The efficient computation of the deepest point or composite depth (cf. Remark \ref{rem:depth median computation} of Part I) is yet more difficult, and recent advances in nonconvex min-max  optimization \citep{Raza2020} may shed new light on the topic.    

The proposed computational inference tool  caters to machine learning applications beyond the standard likelihood setup. For example, given a feedforward neural network, it can be used to evaluate the reliability of a given estimate, or  an event concerned with some properties of the unknowns, which only requires the gradient information that
can be obtained from back propagation with ease.
Moreover, the influence-driven deepest estimation provides a universal means of    accommodating distortions  and anomalies given any criterion or estimation equations.  We hope that the work is helpful to advance the practice of data depth in sophisticated  setups and in higher dimensions.


%
%


{
\bibliographystyle{apalike}
\bibliography{DataDepth}

\begin{thebibliography}{}

\bibitem[Alfons et~al., 2013]{alfons2013sparse}
Alfons, A., Croux, C., and Gelper, S. (2013).
\newblock Sparse least trimmed squares regression for analyzing
  high-dimensional large data sets.
\newblock {\em The Annals of Applied Statistics}, 7(1):226--248.

\bibitem[Anderson, 1951]{And51}
Anderson, T.~W. (1951).
\newblock Estimating linear restrictions on regression coefficients for
  multivariate normal distributions.
\newblock {\em Annals of Mathematical Statistics}, 22:327--351.

\bibitem[Bern and Eppstein, 2002]{Bern2002}
Bern and Eppstein (2002).
\newblock Multivariate regression depth.
\newblock {\em Discrete {\&} Computational Geometry}, 28(1):1--17.

\bibitem[Bijral et~al., 2007]{bijral2007}
Bijral, A.~S., Breitenbach, M., and Grudic, G. (2007).
\newblock Mixture of {Watson} distributions: A generative model for
  hyperspherical embeddings.
\newblock In {\em Proceedings of the Eleventh International Conference on
  Artificial Intelligence and Statistics}, pages 35--42, San Juan, Puerto Rico.

\bibitem[Boothby, 1986]{boothby1986introduction}
Boothby, W.~M. (1986).
\newblock {\em An introduction to differentiable manifolds and Riemannian
  geometry}.
\newblock Academic Press, Orlando, FL.

\bibitem[Boyd and Vandenberghe, 2004]{Boyd2004}
Boyd, S. and Vandenberghe, L. (2004).
\newblock {\em Convex optimization}.
\newblock Cambridge University Press, New York, NY.

\bibitem[Byrd et~al., 1995]{Byrd1995}
Byrd, R., Lu, P., Nocedal, J., and Zhu, C. (1995).
\newblock A limited memory algorithm for bound constrained optimization.
\newblock {\em SIAM Journal on Scientific Computing}, 16(5):1190--1208.

\bibitem[Cai et~al., 2009]{Cai2009}
Cai, T.~T., Xu, G., and Zhang, J. (2009).
\newblock On recovery of sparse signals via $\ell _{1}$ minimization.
\newblock {\em IEEE Transactions on Information Theory}, 55(7):3388--3397.

\bibitem[Dhillon et~al., 2003]{dhillon2003}
Dhillon, I.~S., Marcotte, E.~M., and Roshan, U. (2003).
\newblock Diametrical clustering for identifying anti-correlated gene clusters.
\newblock {\em Bioinformatics}, 19(13):1612--1619.

\bibitem[Edelman et~al., 1998]{Alan1998}
Edelman, A., Arias, T.~A., and Smith, S.~T. (1998).
\newblock The geometry of algorithms with orthogonality constraints.
\newblock {\em SIAM Journal on Matrix Analysis and Applications},
  20(2):303--353.

\bibitem[Fan and Li, 2001]{Fan2001}
Fan, J. and Li, R. (2001).
\newblock Variable selection via nonconcave penalized likelihood and its oracle
  properties.
\newblock {\em Journal of the American Statistical Association},
  96(456):1348--1360.

\bibitem[Harrison and Rubinfeld, 1978]{Harrub1978}
Harrison, D. and Rubinfeld, D. (1978).
\newblock Hedonic housing prices and the demand for clean air.
\newblock {\em Journal of Environmental Economics and Management}, 5:81--102.

\bibitem[Izenman, 2008]{izenbook}
Izenman, A. (2008).
\newblock {\em Modern Multivariate Statistical Techniques: Regression,
  Classification and Manifold Learning}.
\newblock Springer, New York.

\bibitem[Liu and Singh, 1992]{Regina1992}
Liu, R.~Y. and Singh, K. (1992).
\newblock Ordering directional data: concepts of data depth on circles and
  spheres.
\newblock {\em The Annals of Statistics}, 20(3):1468--1484.

\bibitem[L\"{u}tkepohl, 2007]{Ltkepohl2007}
L\"{u}tkepohl, H. (2007).
\newblock {\em New Introduction to Multiple Time Series Analysis}.
\newblock Springer-Verlag Berlin Heidelberg.

\bibitem[Mardia and Jupp, 1999]{mardia1999directional}
Mardia, K.~V. and Jupp, P.~E. (1999).
\newblock {\em Directional statistics}.
\newblock John Wiley \& Sons, Hoboken, NJ.

\bibitem[Mizera, 2002]{mizera2002}
Mizera, I. (2002).
\newblock On depth and deep points: a calculus.
\newblock {\em Ann. Statist.}, 30(6):1681--1736.

\bibitem[Parikh and Boyd, 2014]{Parikh2014}
Parikh, N. and Boyd, S. (2014).
\newblock {Proximal algorithms}.
\newblock {\em Foundations and Trends in Optimization}, 1(3):127--239.

\bibitem[Razaviyayn et~al., 2020]{Raza2020}
Razaviyayn, M., Huang, T., Lu, S., Nouiehed, M., Sanjabi, M., and Hong, M.
  (2020).
\newblock Nonconvex min-max optimization: Applications, challenges, and recent
  theoretical advances.
\newblock {\em IEEE Signal Processing Magazine}, 37(5):55--66.

\bibitem[Reinsel and Velu, 1998]{ReinVelu}
Reinsel, G. and Velu, R. (1998).
\newblock {\em Multivariate Reduced-Rank Regression: Theory and Applications}.
\newblock Springer, New York.

\bibitem[Rothman et~al., 2010]{Rothman2010}
Rothman, A.~J., Levina, E., and Zhu, J. (2010).
\newblock Sparse multivariate regression with covariance estimation.
\newblock {\em Journal of Computational and Graphical Statistics},
  19(4):947--962.

\bibitem[Rousseeuw and Hubert, 1999]{Rousseeuw1999regression}
Rousseeuw, P.~J. and Hubert, M. (1999).
\newblock Regression depth.
\newblock {\em Journal of the American Statistical Association},
  94(446):388--402.

\bibitem[She, 2012]{She2012}
She, Y. (2012).
\newblock An iterative algorithm for fitting nonconvex penalized generalized
  linear models with grouped predictors.
\newblock {\em Computational Statistics \& Data Analysis}, 56(10):2976--2990.

\bibitem[She, 2013]{SheTISPMat}
She, Y. (2013).
\newblock Reduced rank vector generalized linear models for feature extraction.
\newblock {\em Statistics and Its Interface}, 6:197--209.

\bibitem[She, 2016]{she2016c}
She, Y. (2016).
\newblock On the finite-sample analysis of {$\Theta$}-estimators.
\newblock {\em Electron. J. Statist.}, 10(2):1874--1895.

\bibitem[She, 2017]{She2017}
She, Y. (2017).
\newblock Selective factor extraction in high dimensions.
\newblock {\em Biometrika}, 104(1):97--110.

\bibitem[She and Chen, 2017]{She2017RRRR}
She, Y. and Chen, K. (2017).
\newblock Robust reduced-rank regression.
\newblock {\em Biometrika}, 104(3):633--647.

\bibitem[She et~al., 2016]{She2016b}
She, Y., Li, S., and Wu, D. (2016).
\newblock Robust orthogonal complement principal component analysis.
\newblock {\em Journal of the American Statistical Association},
  111(514):763--771.

\bibitem[She et~al., 2022a]{SheDepthI2021}
She, Y., Tang, S., and Liu, L. (2022a).
\newblock {On Generalization and Computation of {T}ukey's Depth: Part I}.
\newblock {\em {Journal of Data Science, Statistics, and Visualisation}}, 2(1).
\newblock DOI:10.52933/jdssv.v2i1.23.

\bibitem[She and Tran, 2019]{SheCV}
She, Y. and Tran, H. (2019).
\newblock On cross-validation for sparse reduced rank regression.
\newblock {\em Journal of the Royal Statistical Society: Series B},
  81:145--161.

\bibitem[She et~al., 2021]{SheetalBreg}
She, Y., Wang, Z., and Jin, J. (2021).
\newblock {Analysis of Generalized Bregman Surrogate Algorithms for Nonsmooth
  Nonconvex Statistical Learning}.
\newblock {\em The Annals of Statistics}, 49(6):3434--3459.

\bibitem[She et~al., 2022b]{SheetalPIQ}
She, Y., Wang, Z., and Shen, J. (2022b).
\newblock { Gaining Outlier Resistance with Progressive Quantiles: Fast
  Algorithms and Theoretical Studies}.
\newblock {\em Journal of the American Statistical Association}.
\newblock To appear.

\bibitem[Sherwood and Maidman, 2019]{rqPen19}
Sherwood, B. and Maidman, A. (2019).
\newblock {\em rqPen: Penalized Quantile Regression}.
\newblock R package version 2.1.

\bibitem[Sra and Karp, 2013]{SRA2013}
Sra, S. and Karp, D. (2013).
\newblock The multivariate watson distribution: Maximum-likelihood estimation
  and other aspects.
\newblock {\em Journal of Multivariate Analysis}, 114:256 -- 269.

\bibitem[Tibshirani, 1996]{Tibshirani1996}
Tibshirani, R. (1996).
\newblock Regression shrinkage and selection via the lasso.
\newblock {\em Journal of the Royal Statistical Society. Series B
  (Methodological)}, 58(1):267--288.

\bibitem[Tukey, 1975]{tukey1975mathematics}
Tukey, J.~W. (1975).
\newblock Mathematics and the picturing of data.
\newblock In {\em Proceedings of the international congress of mathematicians},
  volume~2.

\bibitem[Watson, 1965]{Watson1965}
Watson, G.~S. (1965).
\newblock Equatorial distributions on a sphere.
\newblock {\em Biometrika}, 52(1/2):193--201.

\bibitem[Yang and Zhao, 2020]{packageRRRR}
Yang, Y. and Zhao, Z. (2020).
\newblock {\em RRRR: Online Robust Reduced-Rank Regression Estimation}.

\bibitem[Zhang, 2010]{Zhang2010MCP}
Zhang, C.-H. (2010).
\newblock {Nearly unbiased variable selection under minimax concave penalty}.
\newblock {\em The Annals of Statistics}, 38(2):894--942.

\bibitem[Zhao and Palomar, 2017]{zhao2017robust}
Zhao, Z. and Palomar, D.~P. (2017).
\newblock Robust maximum likelihood estimation of sparse vector error
  correction model.
\newblock In {\em 2017 IEEE Global Conference on Signal and Information
  Processing (GlobalSIP)}, pages 913--917. IEEE.

\end{thebibliography}
}


\end{document}